\documentclass[pra,twocolumn,aps,showpacs,nofootinbib]{revtex4-1}

\usepackage{physics}
\usepackage{amsmath}
\usepackage{amssymb}
\usepackage{graphicx}
\usepackage{float}
\graphicspath{{PNG/}}
\usepackage{xcolor}
\usepackage{multirow}
\usepackage[utf8]{inputenc}

\makeatletter
\newcommand*\bigcdot{\mathpalette\bigcdot@{.5}}
\newcommand*\bigcdot@[2]{\mathbin{\vcenter{\hbox{\scalebox{#2}{$\m@th#1\bullet$}}}}}
\makeatother

\begin{document}
\title{Quench dynamics of mass-imbalanced three-body fermionic systems in a spherical trap}

\author{A. D. Kerin}
\author{A. M. Martin}

\affiliation{School of Physics, University of Melbourne, Parkville, VIC 3010, Australia}

\date{\today}

\begin{abstract}
We consider a system of two identical fermions of general mass interacting with a third distinguishable particle via a contact interaction within an isotropic three-dimensional harmonic trap. We calculate time-dependent observables of the system after it is quenched in s-wave scattering length. To do this we use exact closed form mass-imbalanced hyperspherical solutions to the static three-body problem. These exact solutions enable us to calculate two time-dependent observables, the Ramsey signal and particle separation, after the system undergoes a quench from non-interacting to the unitary regime or vice-versa.
\end{abstract}
\maketitle

\section{Introduction}

Investigating the dynamics of non-equilibrium quantum systems is relevant to many areas in condensed matter physics. Areas of interest include the study of superfluid turbulence, topological states, mesoscopic circuits and can even be extended to neutron star dynamics. In these systems there are many interacting entities, making exact calculations problematic. However, it is possible to theoretically and experimentally examine quantum dynamics when there are only a few quantum bodies in the system of interest. An idealised example of such a system, where few-body quantum dynamics can be studied, is harmonically trapped quantum gases, which can be constructed at the single to few atom limit \cite{serwane2011deterministic, murmann2015two, zurn2013pairing, zurn2012fermionization, PhysRevLett.96.030401}. In this work we focus on the interaction quench dynamics of three fermionic atoms trapped in a spherically symmetric harmonic trap.

We consider a system of two identical fermions which interact via a contact interaction with a third distinguishable particle, which can have a different mass. We consider the scenario where the contact interaction, between the third particle and the two identical fermions, is quenched either from the non-interacting regime to the unitary regime or vice-versa. To do this we utilise the exact solutions for the system \cite{Cui2012,d2018efimov,jonsell2002universal,blume2010breakdown, kerin2022energetics, kestner2007level}. These exact solutions have previously been used to elucidate the thermodynamic properties of quantum gases \cite{PhysRevLett.102.160401, liu2010three, PhysRevLett.96.030401, Cui2012, PhysRevA.85.033634, PhysRevLett.107.030601, mulkerin2012universality, PhysRevA.86.053631, Nature463_2010, Science335_2010,levinsen2017universality, daily2010energy, colussi2018dynamics, colussi2019bunching, enss2022complex}. In the context of quench dynamics the results presented in this paper complement previous studies in two-dimensional \cite{bougas2022dynamical} and one-dimensional systems \cite{pecak2016two,volosniev2017strongly,kehrberger2018quantum, sowinski2019one}.

The paper is structured as follows. In Sec. \ref{sec:Overview} we briefly review the stationary three-body problem of two identical fermions interacting with a third distinguishable particle in a spherically symmetric harmonic trap. In Sec. \ref{sec:Quench} we use the eigenstates of the system to investigate the quench dynamics. In particular we focus on quenches from the non-interacting regime to the strongly interacting (unitary) regime, a forwards quench, and vice-versa, a backwards or reverse quench. For these two quenches we evaluate the Ramsey signal, i.e. the overlap of the time evolving state with the initial state, and post-quench evolution of the particle separation, as defined by the hyperradius. For the Ramsey signal we find that it can be calculated semi-analytically for any initial state for both quenches whilst for the particle separation we find that it can also be calculated semi-analytically for any initial state for the forwards quench but the reverse quench leads to non-physical divergences, as is the case for two-body quench dynamics \cite{kerin2020two}.

\section{Overview of the Three-Body Problem}
\label{sec:Overview}
Our starting point is the Hamiltonian of three non-interacting bodies in a three-dimensional spherical harmonic trap:
\begin{eqnarray}
\hat{H}=\sum_{k=1}^{3} \left[ \frac{-\hbar^2}{2m_{k}}\nabla_{k}^2 +\frac{m_{k} \omega^2 r_{k}^2}{2}\right],
\end{eqnarray}
where $\vec{r}_{k}$ is the position of the $k^{\rm th}$ particle, $m_{k}$ is its mass, and $\omega$ is the trapping frequency.

In this paper we consider the case of two identical fermions interacting with a distinct third particle. We define particle one to be the impurity $(m_{1}=m_{\rm i})$ and particles two and three to be identical $(m=m_{2}=m_{3})$. For convenience we define the reduced mass $\mu=m_{\rm i}m/(m_{\rm i}+m)$, the lengthscale, $a_{\mu}=\sqrt{\hbar/\mu\omega}$, and the mass imbalance $\kappa=m/m_{\rm i}$.

For such a system the hyperspherical formulation \cite{werner2006unitary} gives a closed form solution for the wavefunction. However, because the interactions are enforced with the Bethe-Peierls boundary condition, the wavefunction can only be fully specified in the non-interacting and strongly interacting (unitary) regimes. 

We define the hyperradius $R$ and hyperangle $\alpha$ 
\begin{eqnarray}
R^2=\sqrt{r^2+\rho^2},\quad \alpha=\arctan{(r/\rho)},
\end{eqnarray} 
where
\begin{eqnarray}
\vec{r}&=&\vec{r}_{2}-\vec{r}_{1},\\
\vec{\rho}&=&\frac{1}{\gamma}\left(\vec{r}_{3}-\frac{m_{\rm i}\vec{r}_{1}+m\vec{r}_{2}}{m_{\rm i}+m} \right),\\
\gamma&=&\frac{\sqrt{m_{\rm i}(m_{\rm i}+2m)}}{m+m_{\rm i}},
\end{eqnarray}
and the centre-of-mass (COM) coordinate is 
\begin{eqnarray}
\vec{C}&=&\frac{m_{\rm i}\vec{r}_{1}+m\vec{r}_{2}+m\vec{r}_{3}}{m_{\rm i}+2m}.
\end{eqnarray}
The centre-of-mass Hamiltonian is a simple harmonic oscillator (SHO) Hamiltonian of a single particle of mass $M=2m+m_{\rm i}$ and position $\vec{C}$. As such the centre-of-mass wavefunction is a SHO wavefunction. The relative Hamiltonian is given
\begin{eqnarray}
\hat{H}_{\rm rel}&=&\frac{-\hbar^2}{2\mu}\Bigg( \frac{\partial^2}{\partial R} +\frac{1}{R^2\sin(\alpha)\cos(\alpha)}\frac{\partial^2}{\partial \alpha^2}\left(\cos(\alpha)\sin(\alpha)\bigcdot\right)\nonumber\\
&+&\frac{5}{R}\frac{\partial}{\partial R}-\frac{4}{R^2}-\frac{\hat{\Lambda}_{r}^2}{R^2\sin(\alpha)}-\frac{\hat{\Lambda}_{\rho}^2}{R^2\cos(\alpha)} \Bigg)+\frac{\mu\omega^2R^2}{2},\nonumber\\
\end{eqnarray}
where $\hat{\Lambda}_{r}^2$ and $\hat{\Lambda}_{\rho}^2$ are the angular momentum operators in the $\hat{r}$ and $\hat{\rho}$ coordinate systems.

We define the trial wavefunction \cite{werner2008trapped}
\begin{eqnarray}
\psi_{\rm 3b}^{\rm rel}&=&N_{qls}\frac{F_{qs}(R)}{R^2}(1-\hat{P}_{23})\frac{\varphi_{ls}(\alpha)}{\sin(2\alpha)}Y_{lm}(\hat{\rho}),\label{eq:Ansatz}
\end{eqnarray}
where $N_{qls}$ is the normalisation constant, $F_{qs}(R)$ is the hyperradial wavefunction, $\phi_{ls}(\alpha)=(1-\hat{P}_{23})\varphi_{ls}(\alpha)Y_{lm}(\hat{\rho})/\sin(2\alpha)$ is the hyperangular wavefunction, and $\hat{P}_{23}$ is the particle exchange operator which swaps the positions of particles two and three.

Three conditions determine the functional forms of $F_{qs}(R)$ and $\varphi_{ls}(\alpha)$, 
\begin{eqnarray}
\varphi_{ls}\left(\frac{\pi}{2}\right)&=&0,\\
s^2\varphi_{ls}(\alpha)&=&-\varphi_{ls}''(\alpha)+\frac{l(l+1)}{\cos^2(\alpha)}\varphi_{ls}(\alpha),\\
E_{\rm rel}&=&\frac{-\hbar^2}{4\mu}\left(F''(R)+\frac{F'(R)}{R}\right)\nonumber\\
&+&\left(\frac{\hbar^2s^2}{4\mu R^2}+\mu\omega^2 R^2\right)F(R).
\end{eqnarray}
The first is enforced because a divergence at $\alpha=\pi/2$ is non-physical, the second and third come from the Schr{\"o}dinger equation. $l\in \mathbb{Z}_{\geq0}$ is the angular momentum quantum number, $q\in \mathbb{Z}_{\geq0}$ and $s\in \mathbb{R}_{>0}$ are the energy eigenvalues and the energy is given $E_{\rm rel}=(2q+l+s+1)\hbar\omega$. $F_{qs}(R)$ and $\varphi_{ls}(\alpha)$ are given \cite{werner2006unitary, PhysRevA.74.053604, liu2010three}
\begin{eqnarray}
&&F_{qs}(R)=\tilde{R}^s e^{-\tilde{R}^2/2}L_{q}^{s}\left(\tilde{R}^2\right),
\\
&&\varphi_{ls}(\alpha)=\cos^{l+1}(\alpha)\nonumber\\
&&\qquad\qquad\times {}_{2}F_{1}\left(\frac{l+1-s}{2},\frac{l+1+s}{2};l+\frac{3}{2};\cos^2(\alpha)\right),\nonumber\\\label{eq:AngForm}
\end{eqnarray}
where $\tilde{R}=R/a_{\mu}$, $L_{q}^{s}$ is a Laguerre polynomial, and ${}_{2}F_{1}$ is the Gaussian hypergeometric function.

The contact interactions are enforced by the Bethe-Peierls condition \cite{bethe1935quantum}
\begin{eqnarray}
\lim_{r_{ij}\rightarrow0}\left[\frac{d(r_{ij}\Psi)}{dr_{ij}} \frac{1}{r_{ij}\Psi}\right]=\frac{-1}{a_{\rm s}},
\label{eq:BethePeierls}
\end{eqnarray}
where $\Psi$ is the total three-body wavefunction, $r_{ij}=|\vec{r}_{i}-\vec{r}_{j}|$, and $a_{\rm s}$ is the s-wave scattering length. 

In the non-interacting limit ($a_{\rm s}\rightarrow0$) Eq. (\ref{eq:BethePeierls}) implies, for all values of $\kappa=m/m_{\rm i}$,
\begin{eqnarray}
s=
\begin{cases}
2n+4 & l=0\\
2n+l+2 & l>0
\end{cases},
\end{eqnarray}
where $n\in\mathbb{Z}_{\geq0}$. In the unitary limit ($a_{\rm s}\rightarrow \infty$) the Bethe-Peierls boundary condition gives the transcendental equation
\begin{eqnarray}
&&0=\nonumber\\
&&\frac{d\varphi_{ls}}{d\alpha}\bigg |_{\alpha=0}-(-1)^l\frac{(1+\kappa)^2}{\kappa\sqrt{1+2\kappa}}\varphi_{ls}\left(\arctan(\frac{\sqrt{1+2\kappa}}{\kappa})\right).\nonumber\\\label{eq:Transcendental}
\end{eqnarray}
The values of $s$ at unitarity for $l=0$ and a variety of $\kappa$ are given in Table \ref{tab:sEigenvalues}. 

\begin{table}[H]
\center
\begin{tabular}{|c|c|c|c|}
\hline
$l=0$  & $\kappa=0.1$ &$\kappa=1$ & $\kappa=10$\\
\hline
$n$ & \multicolumn{3}{c|}{$s_{nl}$} \\
\hline
 0 & 2.004\dots & 2.166\dots & 3.316\dots \\\cline{1-4}
 1 & 5.817\dots & 5.127\dots & 4.707\dots \\\cline{1-4}
 2 & 6.195\dots & 7.114\dots & 6.747\dots \\\cline{1-4}
 3 & 9.685\dots & 8.832\dots & 8.876\dots \\
\hline
\end{tabular}
\caption{Some three-body $s$-eigenvalues at unitarity with $l=0$ and $\kappa=0.1,1$, and $10$, (heavy impurity, equal mass and light impurity)} to three decimal places.
\label{tab:sEigenvalues}
\end{table}

From this it is possible, in the non-interacting and unitary regimes, to evaluate the eigenenergies and eigenstates of the three-particle system. From this foundation in the following section we utilise these states to evaluate the quench dynamics of this system.

\section{Quench Dynamics}
\label{sec:Quench}
Below we investigate the behaviour of the system after a quench in the s-wave scattering length, $a_{\rm s}$. We are interested in the forwards quench (non-interacting to unitary) and the backwards quench (unitary to non-interacting). Specifically, we calculate the Ramsey signal, $S(t)$, and the particle separation, $\langle \tilde{R}(t) \rangle$. In order to do this we need to calculate various integrals involving the wavefunction. First we have the Jacobian
\begin{eqnarray}
dV=d\vec{r}_{1}d\vec{r}_{2}d\vec{r}_{3}=\frac{R^5}{4}\sin^2(2\alpha)\gamma^3dR d\alpha d\vec{\Omega}_{r}d\vec{\Omega}_{\rho}d\vec{C}. \quad
\end{eqnarray}
We make the definitions
\begin{eqnarray}
&&\bra{F_{qs}(R)}\ket{F_{qs}(R)}=\int_{0}^{\infty} RF_{qs}(R)^{*}F_{qs}(R)dR,\\
&&\bra{\phi_{ls}(\alpha)}\ket{\phi_{ls}(\alpha)}=\nonumber\\
&&\qquad\int\int\int_{0}^{\pi/2} \phi_{ls}(\alpha)^{*} \phi_{ls}(\alpha)2\sin^2(2\alpha) d\alpha d\vec{\Omega}_{r}d\vec{\Omega}_{\rho}.\quad
\end{eqnarray}

In this work we wish to calculate time-varying observables of a post-quench system. To do this we need the time-dependent post-quench wavefunction. The COM wavefunction is independent of $a_{\rm s}$ and so is unaffected by the quench. As such we only need the time-dependent post-quench \textit{relative} wavefunction, it is given
\begin{eqnarray}
\ket{\psi_{\rm 3b}^{\rm rel}(t)}&=&e^{-i\hat{H}_{\rm rel}t/\hbar}\ket{F_{q_{\rm i}s_{\rm i}}\phi_{l_{\rm i}s_{\rm i}}},\nonumber\\
&=&\sum_{q,s}
\bra{F_{qs}\phi_{l_{\rm i}s}}\ket{F_{q_{\rm i}s_{\rm i}}\phi_{l_{\rm i}s_{\rm i}}}e^{-iE_{ql_{\rm i}s}t/\hbar}\ket{F_{qs}\phi_{l_{\rm i}s}}\nonumber,\\
\end{eqnarray}
where quantum numbers with an i subscript are the initial quantum numbers and the summation is over all eigenstates of the post-quench system.

\subsection{Ramsey signal}
\label{sec:Ramsey}

The Ramsey signal is defined as the wavefunction overlap of the pre- and post-quench wavefunctions, it is given
\begin{eqnarray}
S(t)&=&\bra{\Psi_{\rm i}(t)}\ket{\Psi'(t)},\nonumber\\
S(t)&=&\sum_{j}|\bra{\Psi_{\rm i}(0)}\ket{\Psi'_{j}}|^2e^{-i(E_{j}-E_{\rm i})t/\hbar},
\label{eq:RamseyDefn}
\end{eqnarray}
where $\Psi_{\rm i}$ is the pre-quench wavefunction with energy $E_{\rm i}$ and $\Psi'$ is the post-quench wavefunction. To obtain Eq. (\ref{eq:RamseyDefn}) we have inserted a complete set of post-quench eigenstates, $\Psi'_{j}$, with energies $E_{j}$, where the sum over $j$ is a sum over all post-quench eigenstates \cite{kerin2020two}.

Since the COM wavefunction is unaffected by the quench the Ramsey signal is given
\begin{eqnarray}
S(t)&=&\sum_{q,s}|\bra{F_{q_{\rm i}s_{\rm i}}\phi_{l_{\rm i}s_{\rm i}}}\ket{F_{qs}\phi_{l_{\rm i}s}}|^2e^{-i(E_{ql_{\rm i}s}-E_{q_{\rm i}l_{\rm i}s_{\rm i}})t/\hbar},\quad
\label{eq:RamseySignal}
\end{eqnarray}
where indices with subscript i are the eigenvalues of the initial state and the unlabelled indices correspond to the post-quench eigenvalues. Note that $\bra{\phi_{ls}}\ket{\phi_{l'\neq l s'}}=0$, i.e. hyperangular states of different angular momenta are orthogonal.

To evaluate the Ramsey signal we need to evaluate the hyperradial integral $\bra{F_{qs}}\ket{F_{q's'}}$ and the hyperangular integral $\bra{\phi_{ls}}\ket{\phi_{ls'}}$.

The hyperradial integral is given \cite{srivastava2003remarks}
\begin{widetext}
\begin{eqnarray}
&&\bra{F_{qs}(R)}\ket{F_{q's'}(R)}=\nonumber\\
&&\qquad\frac{a_{\mu}^2}{2}\binom{q+s^{*}}{q}\binom{q'+\frac{s'-s^*}{2}-1}{q'}\Gamma\left(\frac{s^{*}+s'+2}{2}\right)
{}_{3}F_{2}\bigg(-q,\frac{s^{*}+s'+2}{2},\frac{s^{*}-s'+2}{2};s^{*}+1,
\frac{s^{*}-s'+2}{2}-q';1\bigg).\qquad\label{eq:HyperradialInt}
\end{eqnarray}
\end{widetext}

For $s=s'$ and $q\neq q'$ the hyperradial integral vanishes. Evaluating the hyperangular integral is not as straightforward as for the hyperradial integral due to the permutation operator. The hyperangle $\alpha$ is defined in terms of the Jacobi vectors $\vec{r}$ and $\vec{\rho}$, which are defined in terms of $\vec{r}_{1},\vec{r}_{2}$, and $\vec{r}_{3}$. We have defined $\vec{r}$ as being between particles one and two but it is possible to define $\vec{r}$ between any pair of particles. There are then three possible ways to define the Jacobi vectors for the three-body system, these Jacobi sets are related to one another by a ``kinematic rotation'' \cite{nielsen2001three}, a coordinate transform in other words. We perform the hyperangular integral by taking advantage of these relations and ``rotating'' the terms acted upon by $\hat{P}_{23}$ into the same Jacobi set as the term not acted upon by $\hat{P}_{23}$ \cite{fedorov1993efimov, fedorov2001regularization,  braaten2006universality,thogersen2009universality}. However this restricts us to the $l=0$ case due to the presence of the spherical harmonic term making the coordinate transform intractable for $l>0$ as the hyperangular part of the wavefunction becomes a function of $\hat{\rho}$ in addition to $\alpha$. For the general mass case the overlap of $\phi_{0s}$ and $\phi_{0s'}$ is given,
\begin{widetext}
\begin{eqnarray}
&&\bra{\phi_{0s}(\alpha)}\ket{\phi_{0s'}(\alpha)}=8\pi\int_{0}^{\pi/2}\left((1-\hat{P}_{23})\frac{\varphi_{0s}(\alpha)}{\sin(2\alpha)} \right)^{*}\left((1-\hat{P}_{23})\frac{\varphi_{0s'}(\alpha)}{\sin(2\alpha)} \right)\sin^2(2\alpha)d\alpha,\\
&&=16\pi\Bigg[ \int_{0}^{\pi/2}\varphi_{0s}^{*}(\alpha)\varphi_{0s'}(\alpha)d\alpha-\frac{(1+\kappa)^2}{2\kappa\sqrt{1+2\kappa}} \int_{0}^{\pi/2}\varphi_{0s}^{*}(\alpha) \Bigg[  \int_{|\arctan(\sqrt{\frac{1}{\kappa}(2+\frac{1}{\kappa})})-\alpha|}^{\pi/2-|\pi/2-\arctan(\sqrt{\frac{1}{\kappa}(2+\frac{1}{\kappa})})-\alpha|}\varphi_{0s'}(\alpha')d\alpha'\Bigg]d\alpha \Bigg].\:\: \label{eq:HyperangularInt}
\end{eqnarray} 
\end{widetext}
For $l=0$ Eq. (\ref{eq:AngForm}) reduces to \cite{werner2008trapped, fedorov2001regularization}
\begin{eqnarray}
\varphi_{0s}(\alpha)\propto\sin(s\left(\frac{\pi}{2}-\alpha\right)).\quad
\label{eq:AngWavefunc}
\end{eqnarray}

Note that combining Eqs. (\ref{eq:HyperangularInt}) and (\ref{eq:AngWavefunc}) for $\kappa=1$ (the equal mass case) does not give the same result as in Refs. \cite{colussi2019two, werner2008trapped} and only agrees when $s=s'$ is one of the solutions to Eq. (\ref{eq:Transcendental}) for $\kappa=1$. This is because the latter references appropriately substitute Eq. (\ref{eq:Transcendental}) for $\kappa=1$ into the result for the hyperangular integral to simplify the expression. 

However, for a general value of $\kappa$ we have to be more careful. For a specific $\kappa$ the $s$ eigenvalues produced by Eq. (\ref{eq:Transcendental}) are orthogonal to one another when Eq. (\ref{eq:HyperangularInt}) is evaluated only for that specific value of $\kappa$, i.e. the $s$-eigenspectrum for $\kappa=x$ produces an orthogonal set of states only when Eq. (\ref{eq:HyperangularInt}) is evaluated using $\kappa=x$. However, the non-interacting values of $s$ ($s=4,6,8,\dots$) are orthogonal to one another whatever the value of $\kappa$. 

\begin{figure}
\includegraphics[height=4.5cm, width=8.5cm]{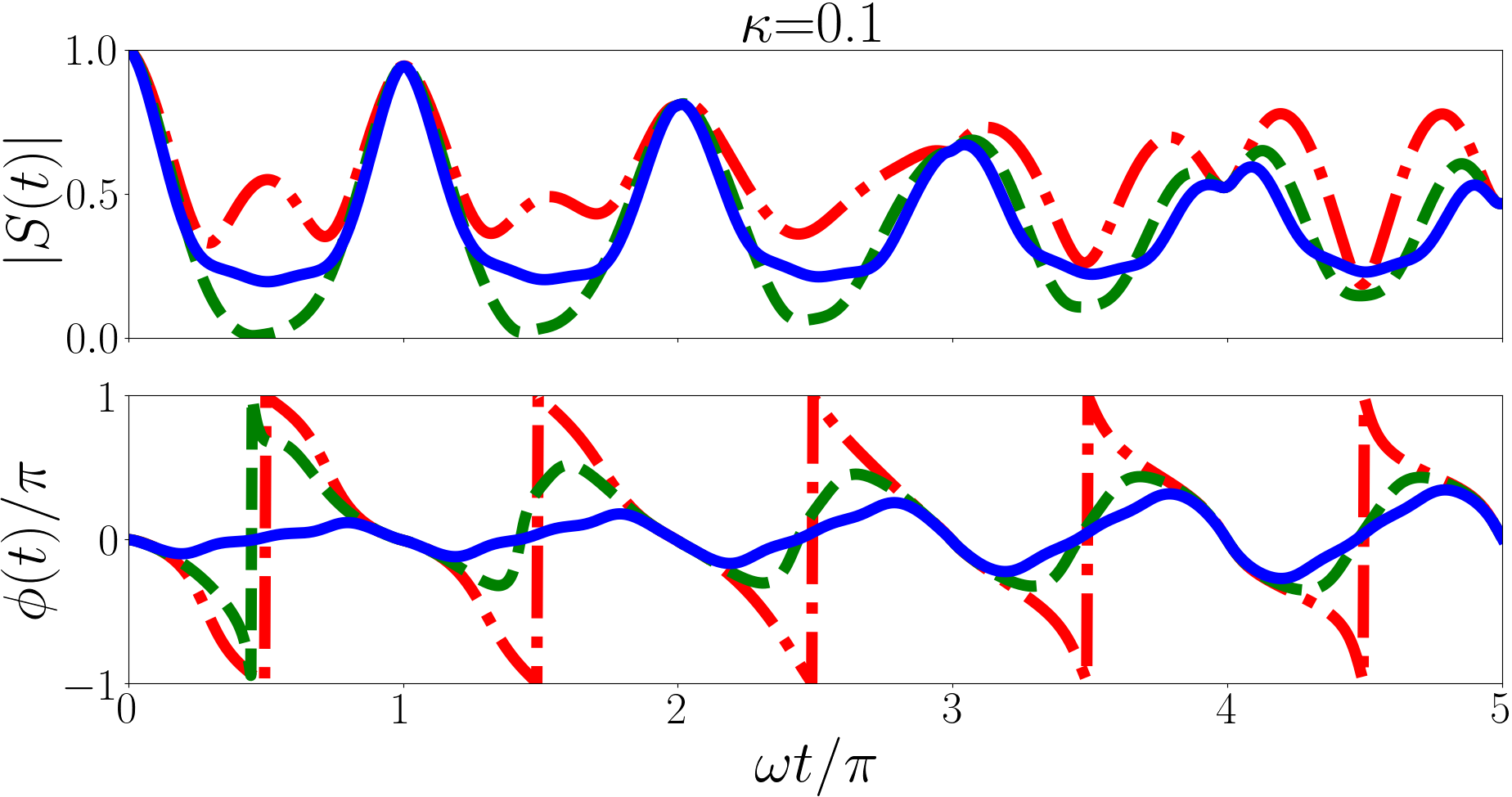}
\includegraphics[height=4.5cm, width=8.5cm]{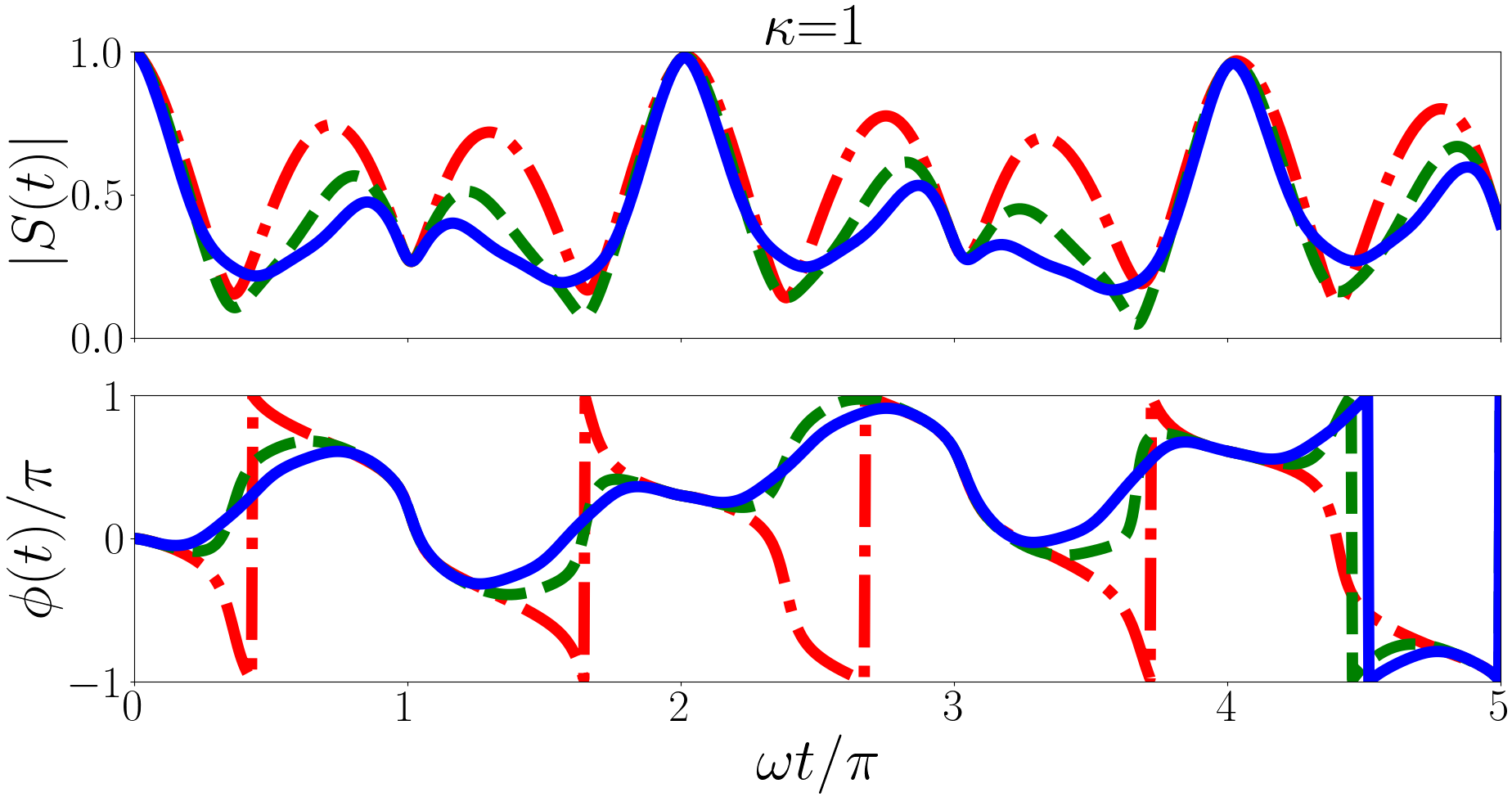}
\includegraphics[height=4.5cm, width=8.5cm]{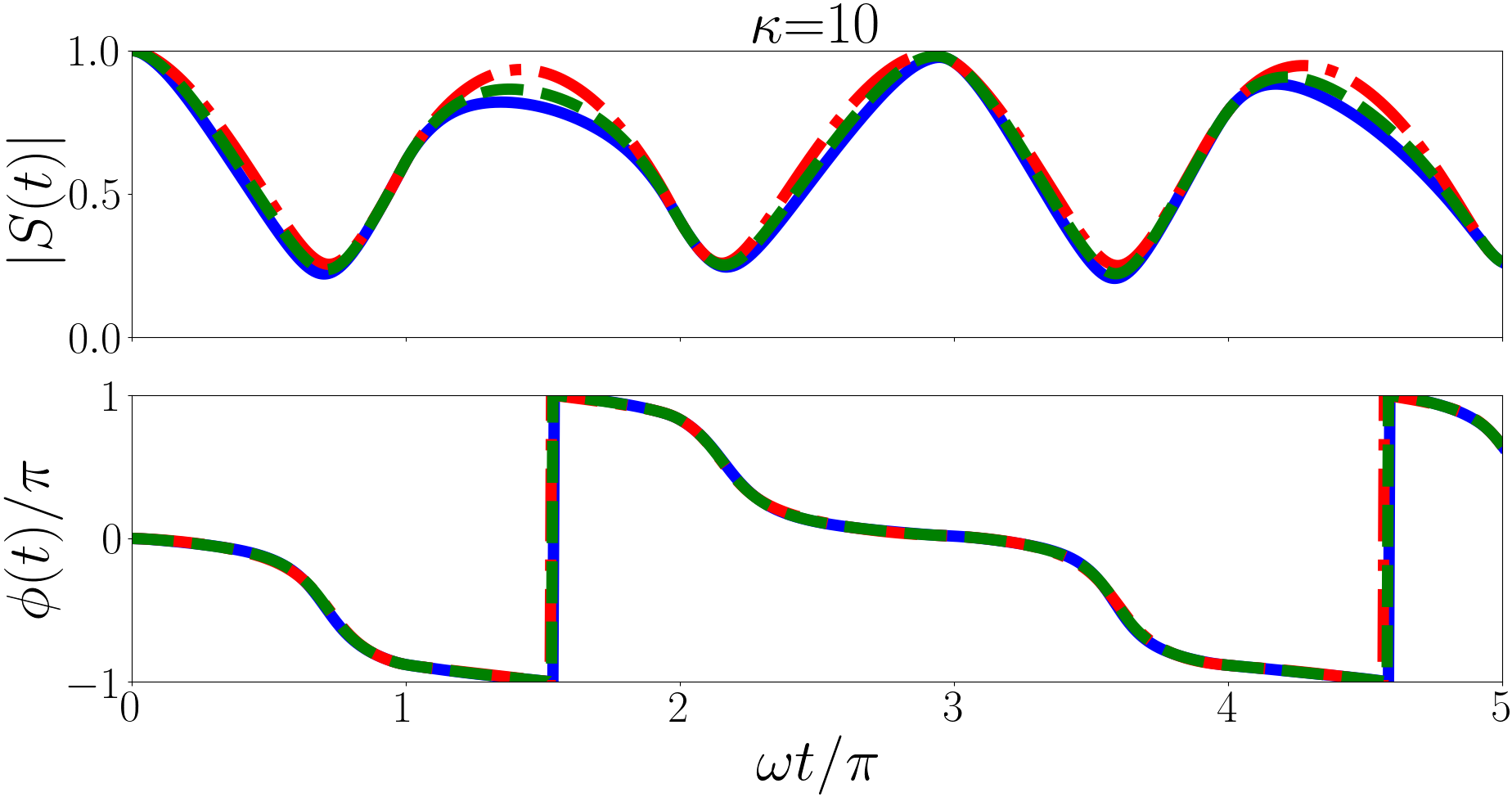}
\caption{Ramsey signal for the forwards quench for a variety initial states and mass imbalances. In all panels the dot-dashed red line corresponds to $(q_{\rm i},s_{\rm i})=(0,4)$, the dashed green line corresponds to $(q_{\rm i},s_{\rm i})=(1,4)$, and the solid blue line corresponds to $(q_{\rm i},s_{\rm i})=(2,4)$. The upper, middle, and lower plots correspond to $\kappa=0.1,1$, and $10$ (heavy impurity, equal mass and light impurity) respectively. Each Ramsey signal is obtained by evaluating Eq. (\ref{eq:RamseySignal}) with 10 terms in each sum, i.e. 100 terms total. In this limit we find that the sum is convergent.}
\label{fig:RamseyForwards}
\end{figure}

\begin{figure}
\includegraphics[height=4.5cm, width=8.5cm]{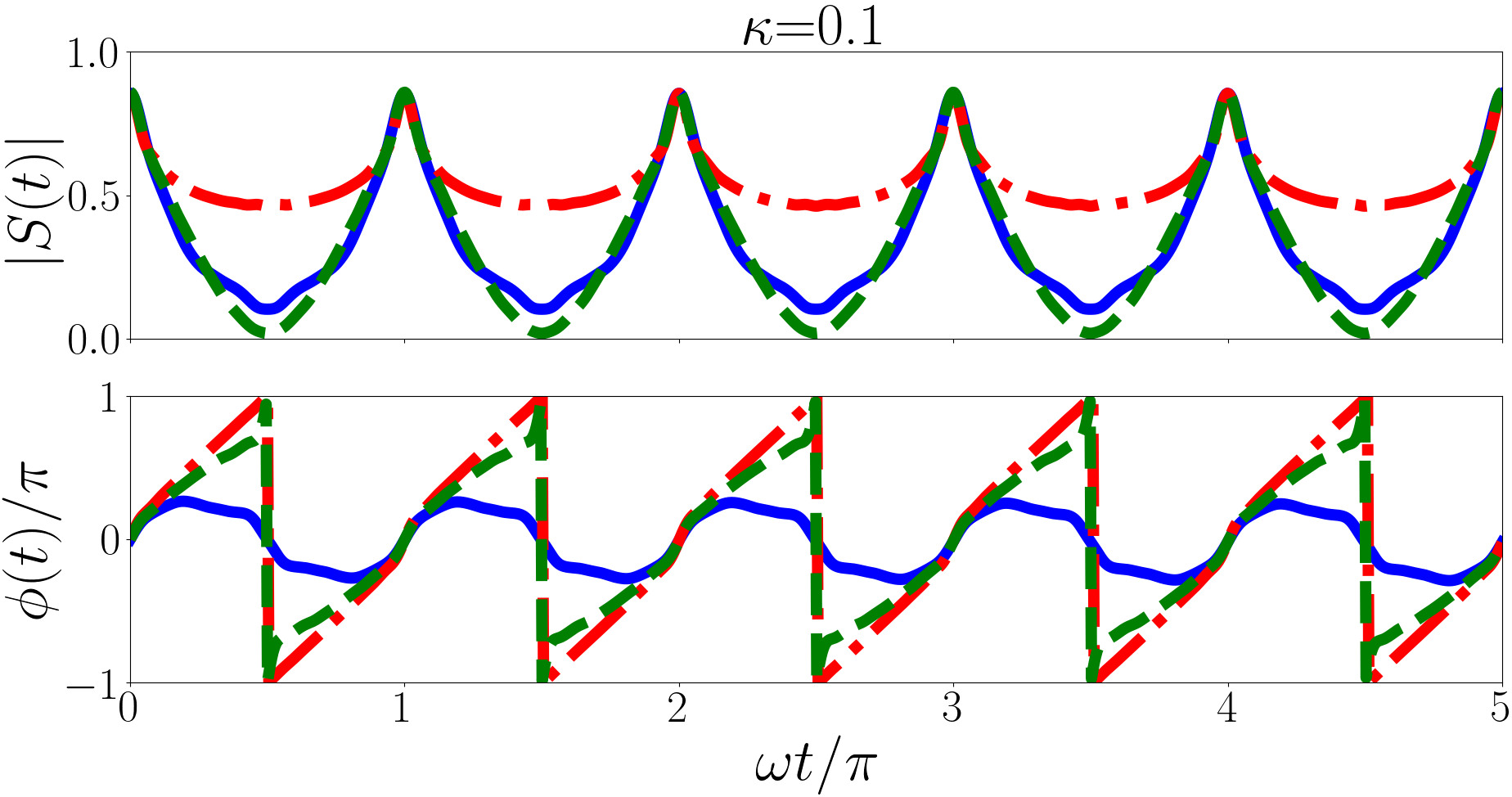}
\includegraphics[height=4.5cm, width=8.5cm]{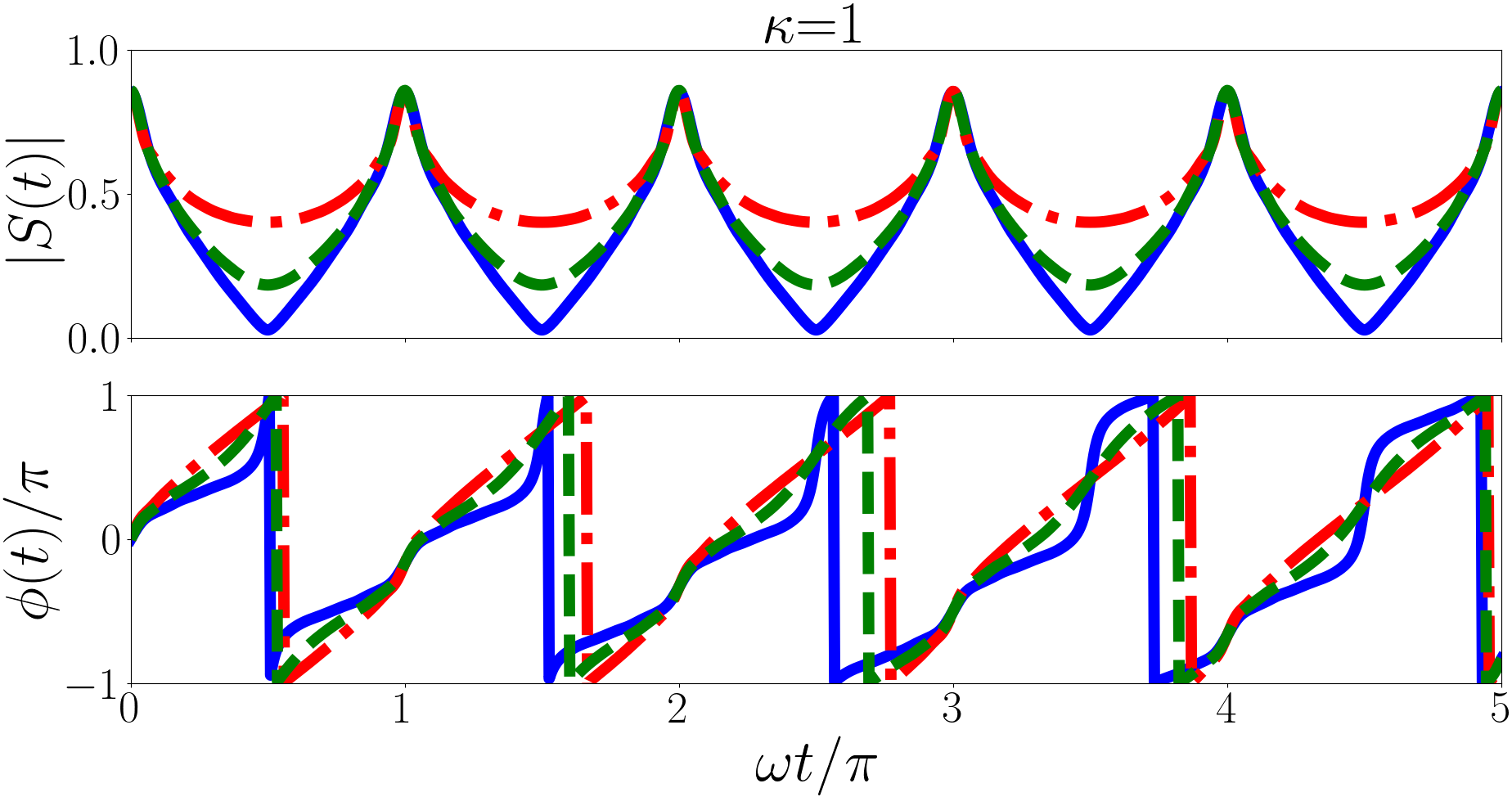}
\includegraphics[height=4.5cm, width=8.5cm]{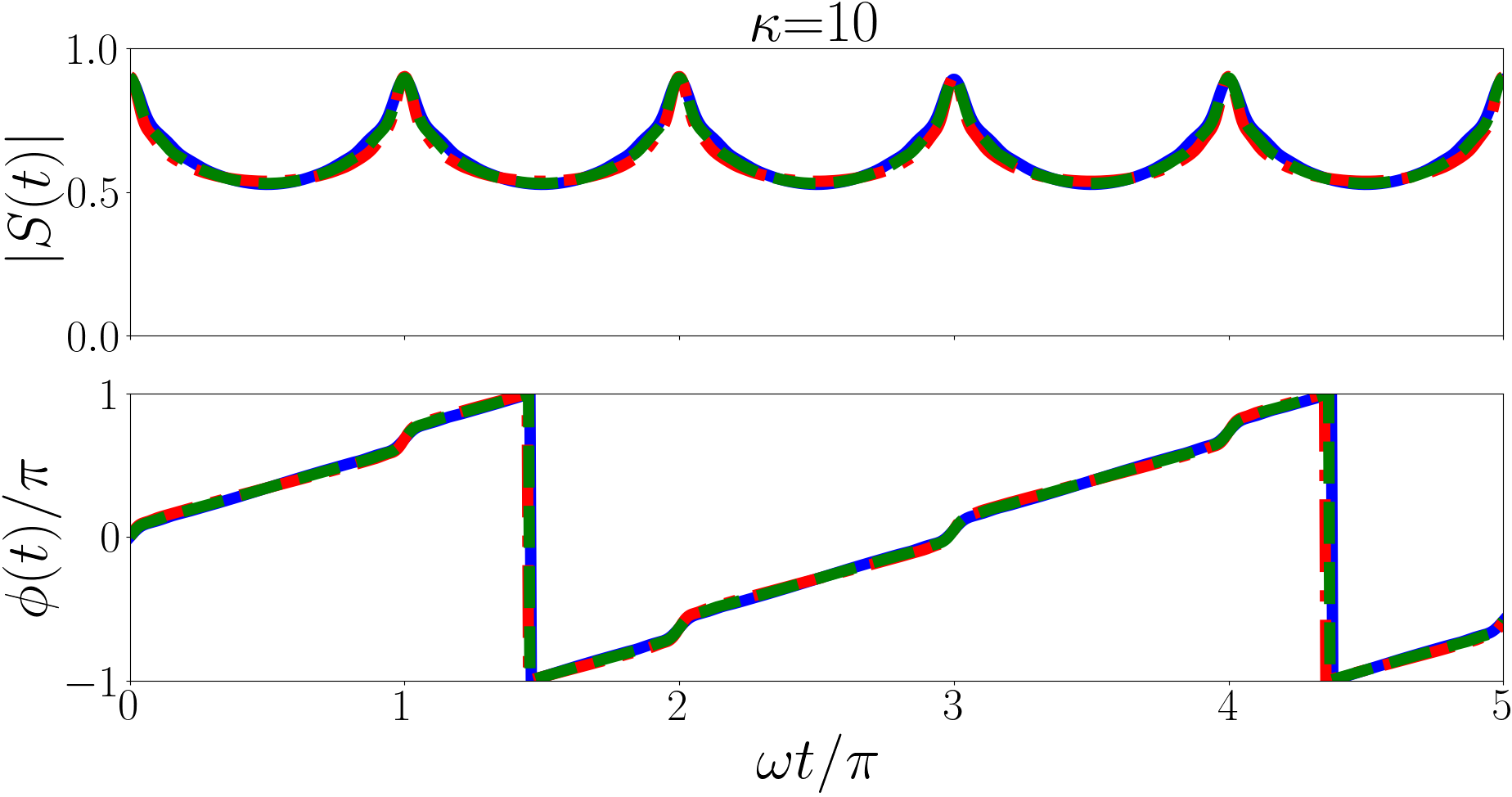}
\caption{Ramsey signal for the backwards quench for a variety of initial states and mass imbalances. In all panels the dot-dashed red line corresponds to $q_{\rm i}=0$, the dashed green line corresponds to $q_{\rm i}=1$, and the solid blue line corresponds to $q_{\rm i}=2$. The upper, middle and lower plots correspond to $\kappa=0.1,1$, and $10$ (heavy impurity, equal mass and light impurity) and $s_{\rm i}=2.004\dots,2.166\dots$, and $3.316\dots$ respectively. Each Ramsey signal is obtained by evaluating Eq. (\ref{eq:RamseySignal}) with 10 terms in each sum, i.e. 100 terms total. In this limit we find that the sum is convergent.}
\label{fig:RamseyBackwards}
\end{figure}

Now that we can calculate the wavefunction overlaps we can evaluate the Ramsey signal for the forwards and backwards quenches for any initial state with $l=0$ and any mass imbalance. In Figs. \ref{fig:RamseyForwards} and \ref{fig:RamseyBackwards} we plot the forwards and backwards quenches respectively for a variety of $\kappa$ and initial states.

Before we discuss the properties of the Ramsey signals, in Figs. \ref{fig:RamseyForwards} and \ref{fig:RamseyBackwards}, it is worth exploring what we may expect in a general sense. Assuming in Eq. (\ref{eq:RamseySignal}) that the sum is dominated by a few terms, the general form for $S(t)$ can be represented by $S(t)\approx A e^{-a it}+B e^{-b it}+C e^{-c i	t}$. The magnitude of the signal oscillates with angular frequencies $(a-b), (b-c), (a-c)$ with the more heavily weighted terms being more significant. However the \textit{phase} of the signal is dominated by the angular frequencies of the most heavily weighted terms not the difference between angular frequencies of different terms. In our case the angular frequencies ($a,b,c,\dots$) are the differences between a post-quench eigenenergy and the initial energy, and the difference between the angular frequencies ($(a-b), (b-c), (a-c),\dots$) is the difference between different post-quench eigenenergies.

In Fig. \ref{fig:RamseyForwards} we plot the Ramsey signals of the forwards quench for a variety of initial states and mass imbalances. We find an irregularly repeating Ramsey signal, this is in contrast to similar calculations performed for the two-body problem \cite{kerin2020two} where the signal is periodic with period $\pi/\omega$. This irregular periodicity arises from the unitary $s$-eigenspectrum. In this case angular frequencies of each term in Eq. (\ref{eq:RamseySignal}) are irrational and so are the differences between them, in general. These irrational angular frequencies lead to the irregular phase and magnitude. 

For the $\kappa=0.1$ (heavy impurity) forwards quench the most heavily weighted terms in Eq. (\ref{eq:RamseySignal}) for the $(q_{\rm i},s_{\rm i})=(0,4)$ initial state (red line in the upper panel of Fig. \ref{fig:RamseyForwards} are the $|\bra{F_{0,4}\phi_{0,4}}\ket{F_{0,2.004}\phi_{0,2.004}}|^2\approx0.539$, $|\bra{F_{0,4}\phi_{0,4}}\ket{F_{0,5.817}\phi_{0,5.817}}|^2\approx0.134$ and $|\bra{F_{0,4}\phi_{0,4}}\ket{F_{1,2.004}\phi_{0,2.004}}|^2\approx0.179$ terms. The corresponding angular frequencies are $\approx -2\omega$, $\approx1.8\omega$, $\approx0.004\omega$ respectively. Hence the phase displays features with a period of $\approx \pi/\omega$, and the magnitude has three main modes with periods $\approx 0.5\pi/\omega$, $\pi/\omega$ and $\approx1.1\pi/\omega$. 

For the $\kappa=1$ (equal mass) forwards quench the most significant terms for the $(q_{\rm i},s_{\rm i})=(0,4)$ initial state are $|\bra{F_{0,4}\phi_{0,4}}\ket{F_{0,2.166}\phi_{0,2.166}}|^2\approx0.499$, $|\bra{F_{0,4}\phi_{0,4}}\ket{F_{0,5.127}\phi_{0,5.127}}|^2\approx0.297$, and $|\bra{F_{0,4}\phi_{0,4}}\ket{F_{1,2.166}\phi_{1,2.166}}|^2\approx0.13$. The corresponding angular frequencies are $\approx -1.8\omega$, $\approx1.1 \omega$ and $\approx0.17\omega$ respectively. The phase is then dominated by a mode with period $\approx1.1\pi/\omega$ and the magnitude is dominated by modes with periods $\approx2\pi/3\omega, \pi/\omega$ and $\approx2\pi/\omega$.

For the $\kappa=10$ (light impurity) forwards quench the most significant terms for the $(q_{\rm i},s_{\rm i})=(0,4)$ initial state are $|\bra{F_{0,4}\phi_{0,4}}\ket{F_{0,3.3169}\phi_{0,3.3169}}|^2\approx0.607$, and $|\bra{F_{0,4}\phi_{0,4}}\ket{F_{0,4.707}\phi_{0,4.707}}|^2\approx0.350$. The corresponding angular frequencies are $\approx-0.7\omega/\pi$ and $\approx0.7\omega/\pi$ respectively hence the phase has an approximate periodicity of $2.8\pi/\omega$ and the magnitude has a period of $\approx 1.4\pi/\omega$.

In Fig. \ref{fig:RamseyBackwards} we plot the Ramsey signals of the backwards quench for a variety of initial states and mass imbalances. We find a strongly periodic magnitude. In the non-interacting regime we have $s=4,6,8,\dots$ for every value of $\kappa$. This means that every oscillatory term in the magnitude has an angular frequency that is a multiple of two, hence the period of $\pi/\omega$. However the behaviour of the phase is still dominated by the largest terms in the summation. For $\kappa=0.1$ (heavy impurity) and $(q_{\rm i}=0, s_{\rm i}=2.00449\dots)$ the largest term is $|\bra{F_{0,2.004}\phi_{0,2.004}}\ket{F_{0,4}\phi_{0,4}}|^2\approx 0.539$ and this defines the period of $\approx\pi/\omega$ we see in the phase. Similarly for $\kappa=1$ (equal mass) and $(q_{\rm i}=0, s_{\rm i}=2.166\dots)$ the largest term is $|\bra{F_{0,2.166}\phi_{0,2.166}}\ket{F_{0,4}\phi_{0,4}}|^2\approx0.499$ defining a period of $\approx 1.1\pi/\omega$ for the phase, and for $\kappa=10$ (light impurity) and $(q_{\rm i}=0, s_{\rm i}=3.3169\dots)$ the largest term is $|\bra{F_{0,3.3169}\phi_{0,3.3169}}\ket{F_{0,4}\phi_{0,4}}|^2\approx 0.607$ defining a period of $\approx2.8\pi/\omega$ for the phase.

\subsection{Particle separation}
\label{sec:Separation}

The Ramsey signal is not the only quench observable we investigate. We can also calculate the expectation value of $\tilde{R}$, i.e. the particle separation. 

The expectation value of $\tilde{R}(t)$ is given
\begin{eqnarray}
&&\langle \tilde{R}(t)\rangle = \bra{\Psi(t)}\tilde{R}\ket{\Psi(t)}\nonumber\\
&& = \sum_{j,j'}\bra{\Psi_{\rm i}(0)}\ket{\Psi'_{j}}\bra{\Psi'_{j'}}\ket{\Psi_{\rm i}(0)} \bra{\Psi'_{j}}\tilde{R}\ket{\Psi'_{j'}} e^{-i(E_{j'}-E_{j})t/\hbar},\nonumber\\
\end{eqnarray}
where $\Psi_{\rm i}$ is the initial pre-quench state with energy $E_{\rm i}$ and $\Psi$ is the post-quench state. $\Psi'_{j}$ and $\Psi'_{j'}$ are eigenstates of the post-quench system with eigenenergy $E_{j}$ and $E_{j'}$ respectively, with the sum over $j$ and $j'$ taken over all post-quench eigenstates.

The COM wavefunction is independent of the interparticle interaction and does not change when the system is quenched and does not impact the post-quench dynamics. Due to the hyperangular wavefunction's orthogonality in $s$, two sums over $s$ and $s'$ collapse into a single sum over $s$. Hence $\langle \tilde{R}(t) \rangle$ is given
\begin{eqnarray}
\langle \tilde{R}(t) \rangle &=& \sum_{q',q}\sum_{s}\bra{F_{q_{\rm i}s_{\rm i}}\phi_{l_{\rm i}s_{\rm i}}}\ket{F_{q's}\phi_{l_{\rm i}s}}\bra{F_{qs}\phi_{l_{\rm i}s}}\ket{F_{q_{\rm i}s_{\rm i}}\phi_{l_{\rm i}s_{\rm i}}}\nonumber\\
&\;& \times \bra{F_{q's}\phi_{l_{\rm i}s}}\tilde{R}\ket{F_{qs}\phi_{l_{\rm i}s}}e^{-i(E_{ql_{\rm i}s}-E_{q'l_{\rm i}s})t/\hbar}.\label{eq:ExpectR}
\end{eqnarray}

All integrals required for calculating Eq. (\ref{eq:ExpectR}) are calculated in Sec. \ref{sec:Ramsey} except for $\bra{F_{q's}}\tilde{R}\ket{F_{q's}}$. This is given \cite{srivastava2003remarks}
\begin{eqnarray}
&&\bra{F_{qs}}\tilde{R}\ket{F_{q's'}}=\nonumber\\
&&\frac{a_{\mu}^2}{2}\binom{q+s}{q}\binom{q'+\frac{s'}{2}-\frac{s}{2}-\frac{3}{2}}{q'}\Gamma\left(\frac{s+s'+3}{2}\right)\nonumber\\
&\;& \times{}_{3}F_{2}\left(-q,\frac{s'+s+3}{2},\frac{s-s'+3}{2};s+1,\frac{s-s'+3}{2}-q';1\right)\nonumber.\\
\end{eqnarray}
This result combined with the results of Sec. \ref{sec:Ramsey} allows us to calculate $\langle \tilde{R}(t) \rangle$ for the forwards and backwards quench for any initial state with $l=0$ and any mass imbalance.

\begin{figure}
\includegraphics[height=4.5cm, width=8.5cm]{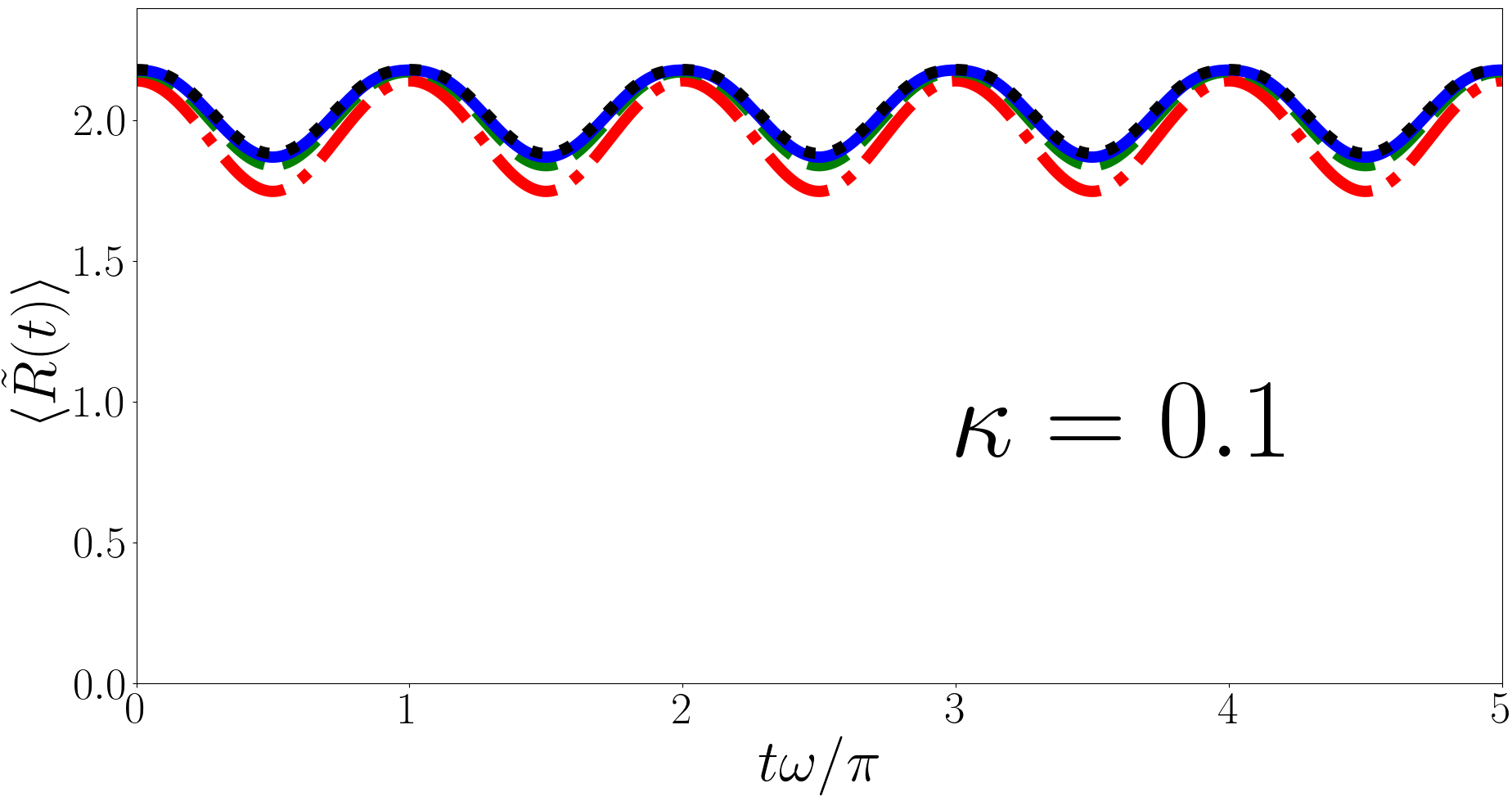}
\includegraphics[height=4.5cm, width=8.5cm]{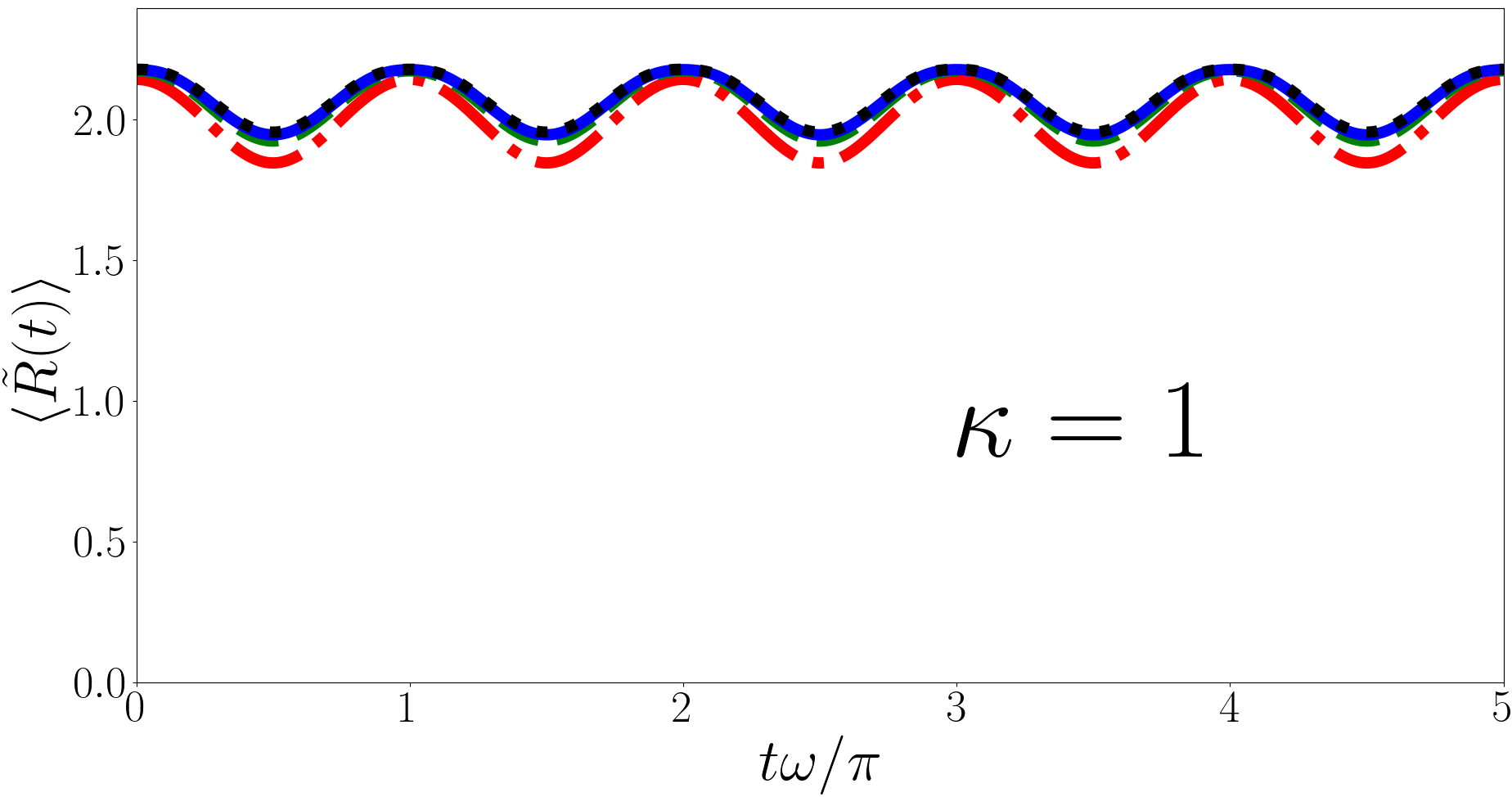}
\includegraphics[height=4.5cm, width=8.5cm]{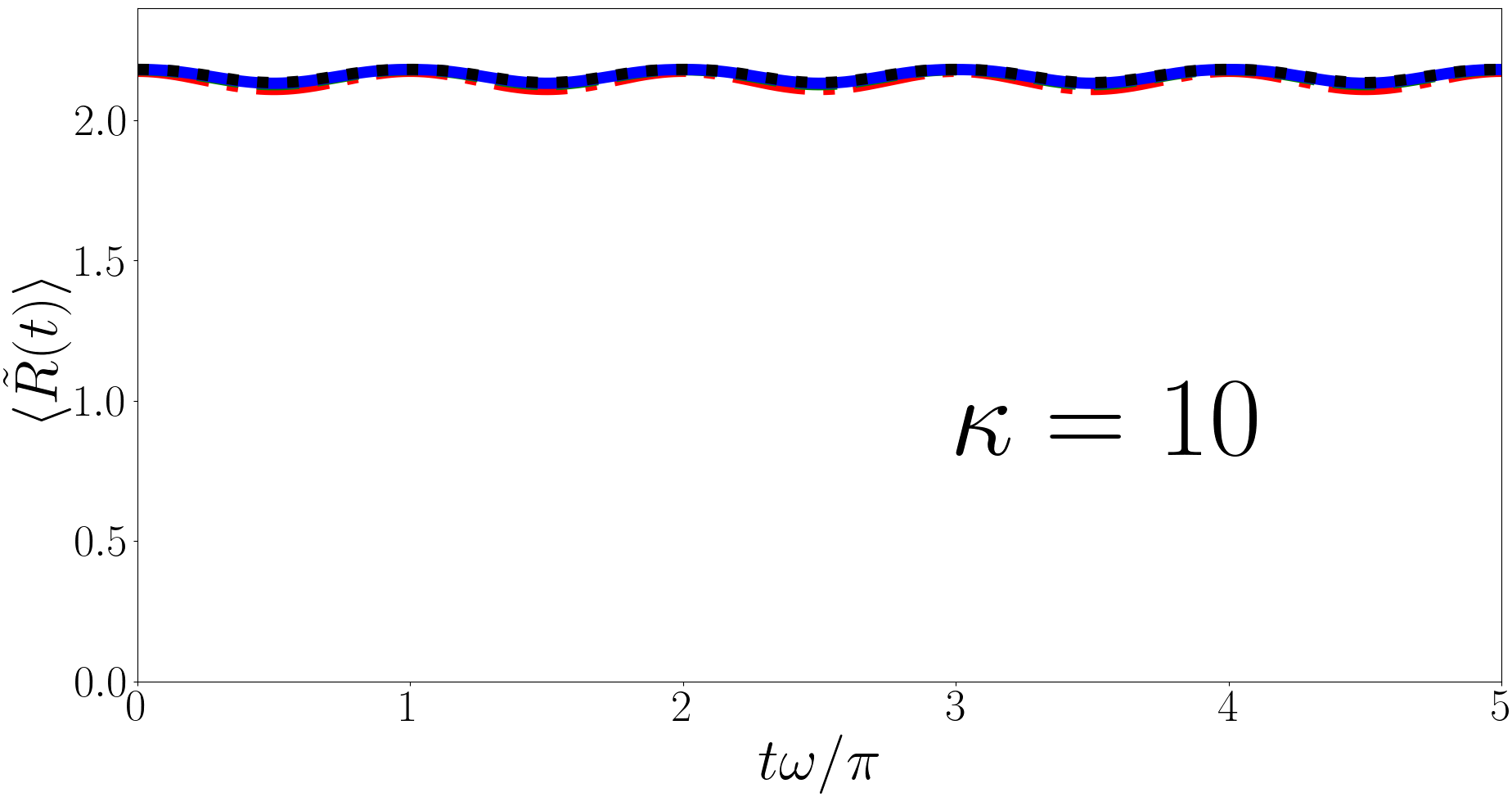}
\caption{$\langle \tilde{R}(t)\rangle$ for the forwards quench from the non-interacting ground state, $(q_{\rm i},s_{\rm i})=(0,4)$, for $\kappa=0.1$ (upper panel, heavy impurity), $\kappa=1$ (middle panel equal mass) and $\kappa=10$ (lower panel light impurity). The dot-dashed red line corresponds to $N_{\rm max}=3$, the dashed green line to $N_{\rm max}=6$, the solid blue to $N_{\rm max}=12$ and the dotted black line to $N_{\rm max}=24$. As can be seen from the plots the sum is convergent with $N_{\rm max}$.}
\label{fig:ExpectRForwards}
\end{figure}

In Sec. \ref{sec:Ramsey} we note that the magnitude and phase of the Ramsey signal for the forwards quench is irregular and this is because the $s$ eigenvalues are irrational at unitarity. Additionally the phase for the backwards quench is also irregular and this is due to the irrationality of $s_{\rm i}$. However the angular frequencies of the terms in Eq. (\ref{eq:ExpectR}) are always even integers because the $s$ contributions to the energies cancel leaving angular frequencies proportional to $(2q'-2q)\omega$ and $q,q'\in\mathbb{Z}_{\geq0}$. This results in the angular frequency of every term in the summation being a multiple of $2\omega$ causing $\langle \tilde{R}(t)\rangle$ to have a period of $\pi/\omega$ in both the forwards and backwards quench.

In Fig. \ref{fig:ExpectRForwards} we have plotted $\langle \tilde{R}(t) \rangle$ for the forwards quench for $\kappa=0.1$, $\kappa=1$ and $\kappa=10$ (heavy impurity, equal mass, and light impurity) in the upper, middle and lower panels respectively. For each $\kappa$ we have taken the initial state to be the ground state and calculated the sum in Eq. (\ref{eq:ExpectR}) over $q$, $q'$ and $s$ up to $N_{\rm max}=3,6,12$ and $24$, where $N_{\rm max}$ is the number of terms in each individual sum, so there are $N_{\rm max}^3$ terms total. In each case we find that the results for $\langle \tilde{R}(t) \rangle$ have converged at $N_{\rm max}=24$. As discussed above $\langle \tilde{R}(t) \rangle$ oscillates with a period $\pi/\omega$.

Additionally we observe that as $\kappa$ increases (impurity becomes lighter) the amplitude of the oscillation decreases. In the $\kappa\rightarrow\infty$ limit the unitary $s$-eigenspectrum approaches $s\rightarrow 4,6,8\dots$ \cite{kerin2022energetics} and the initial non-interacting ground state overlaps perfectly with the unitary ground state. As such the amplitude decreases as $\kappa$ increases until eventually $\langle \tilde{R} \rangle$ reaches a constant value of $2.18\dots$ which is $\langle \tilde{R} \rangle$ for $q=0$, $s=4$. This is also $\langle \tilde{R}(t=0) \rangle$, the maximum particle separation does not change with $\kappa$ but the minimum increases to decrease the amplitude. Conversely in the $\kappa\rightarrow0$ (heavy impurity) limit the unitary $s$-eigenspectrum approaches $s\rightarrow 2,6,10,14\dots$, analytically this presents a problem. The non-interacting ground state, $s_{\rm i}=4$, is orthogonal to the unitary $s$-eigenspectrum because it is a subset of the non-interacting eigenspectrum, except $s=2$ which is forbidden for $l=0$ because it causes the hyperangular part of the wavefunction to be zero. Numeric investigations for $\kappa$ as small as $10^{-3}$ imply that the minimum $\langle \tilde{R}(t) \rangle$ asymptotes to $\approx 1.7\dots$ as $\kappa$ becomes very small. The maximum particle separation remains at $2.18\dots a_{\mu}$ whatever the value of $\kappa$ because the initial non-interacting state does not depend on $\kappa$.

\begin{figure}
\includegraphics[height=4.5cm, width=8.5cm]{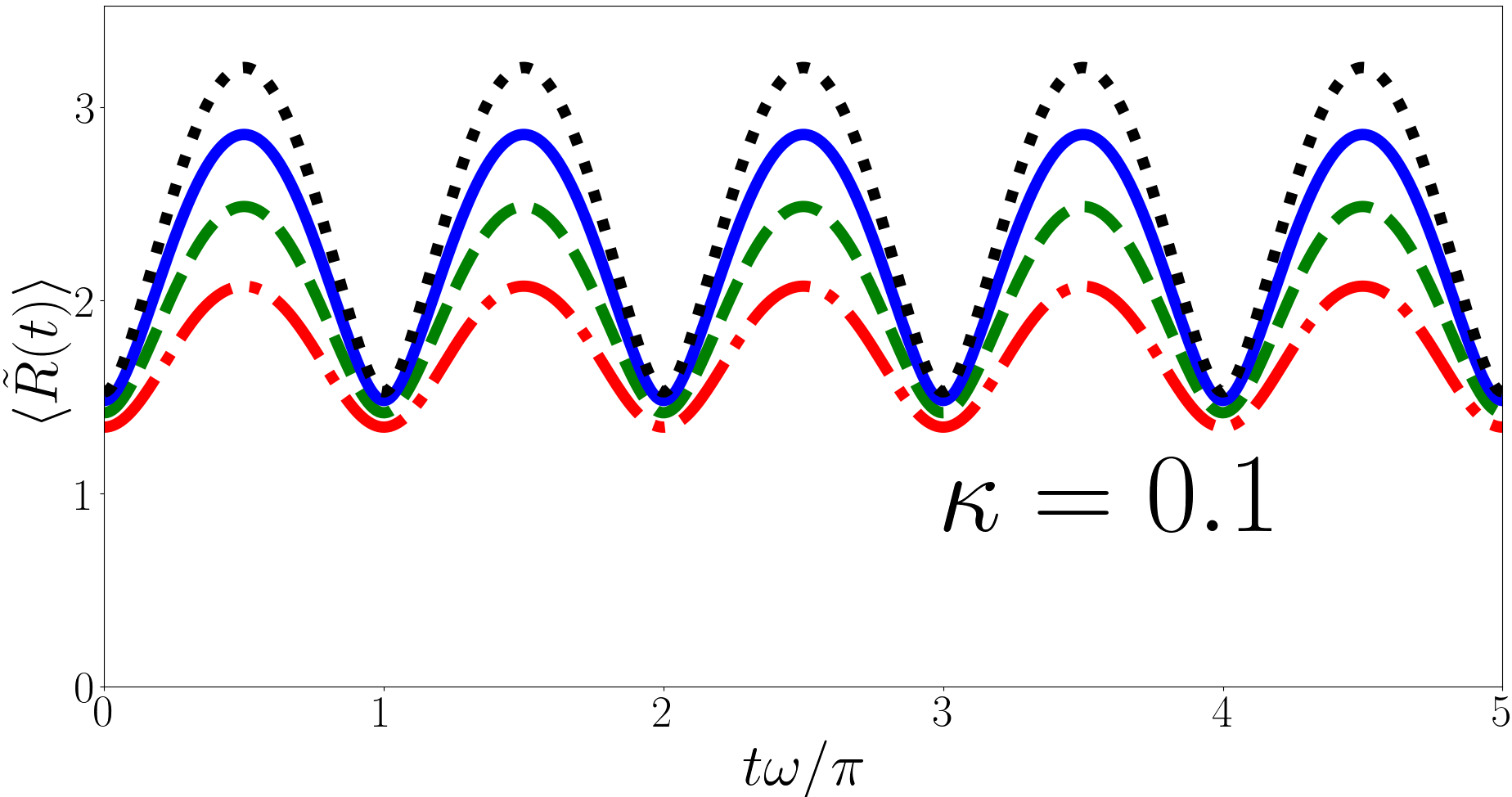}
\includegraphics[height=4.5cm, width=8.5cm]{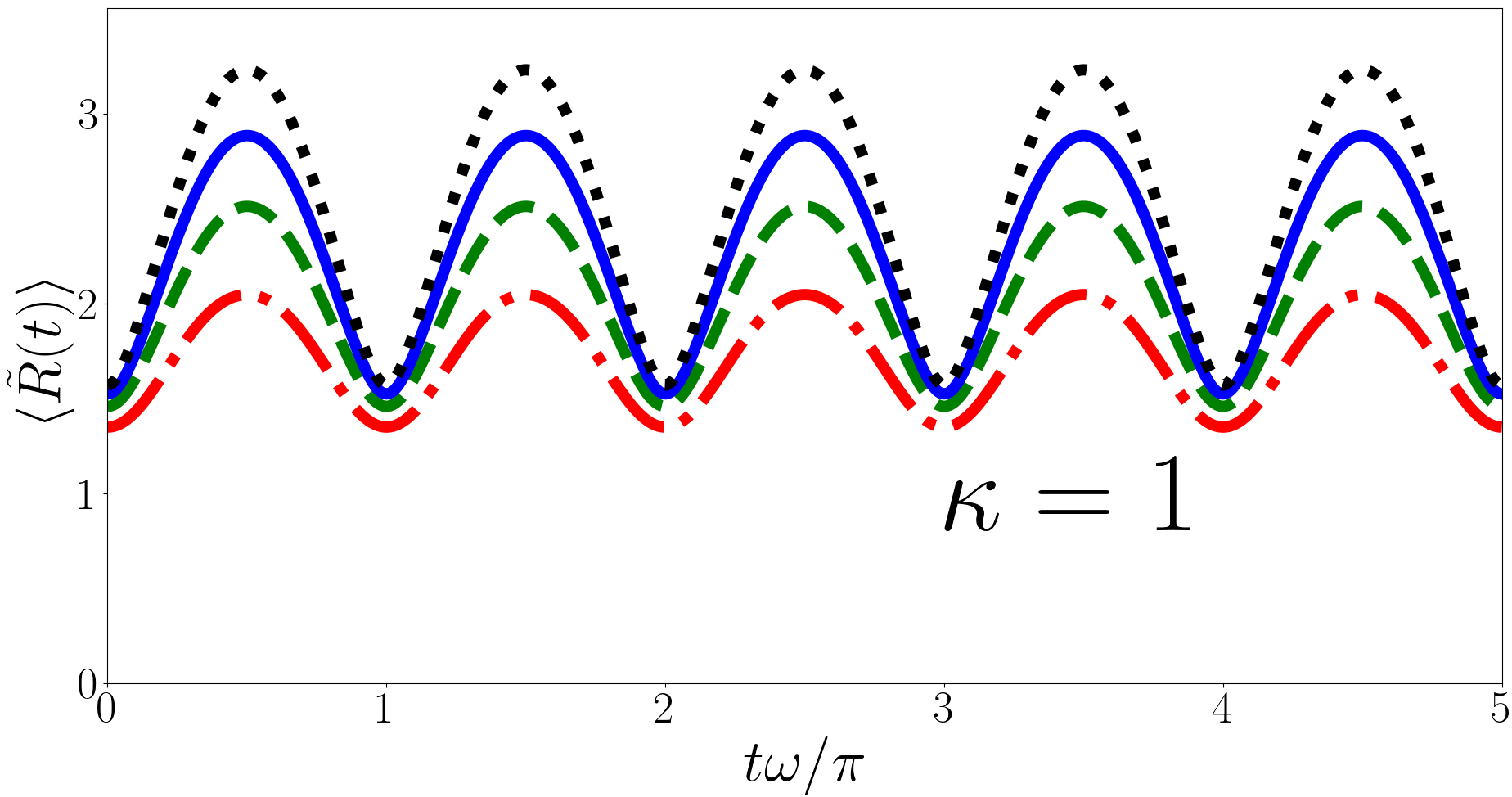}
\includegraphics[height=4.5cm, width=8.5cm]{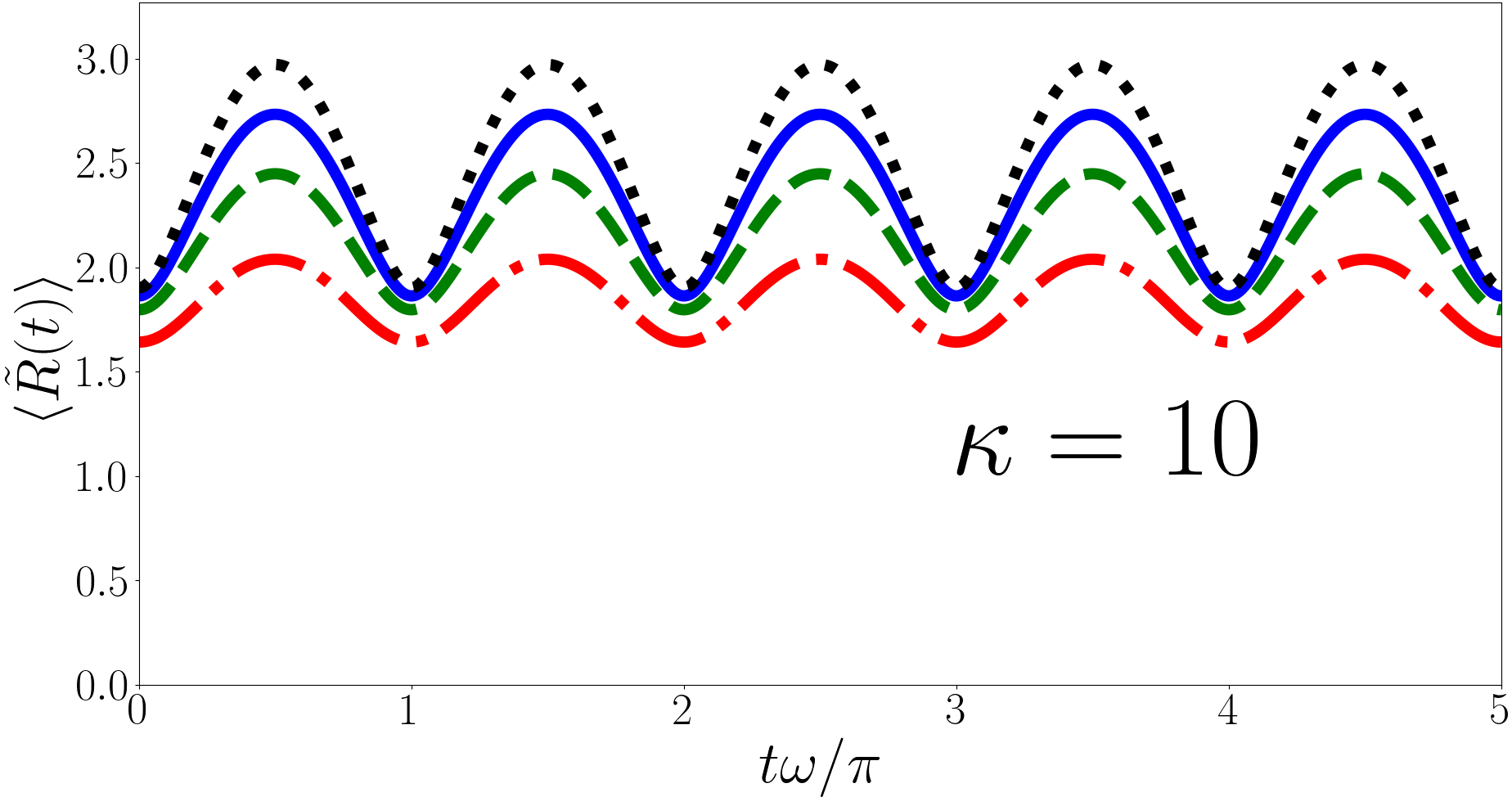}
\caption{$\langle \tilde{R}(t)$ for the backwards quench from the interacting ground state for a variety of $\kappa$. In all panels $q_{\rm i}=0$, in the upper panel ($\kappa=0.1$, heavy impurity) $s_{\rm i}=2.004\dots$, in the middle panel ($\kappa=1$, equal mass) $s_{\rm i}=2.166\dots$, and in the lower panel ($\kappa=10$, light impurity) $s_{\rm i}=3.3169\dots$. The dot-dashed red line corresponds to $N_{\rm max}=3$, the dashed green line to $N_{\rm max}=6$, the solid blue to $N_{\rm max}=12$ and the dotted black line to $N_{\rm max}=24$. This sum diverges logarithmically with $N_{\rm max}$ away from $t=n\pi/\omega$.}
\label{fig:ExpectRBackwards}
\end{figure}

In Fig. \ref{fig:ExpectRBackwards} we have considered $\langle \tilde{R}(t) \rangle$ for the backwards quench for the $\kappa=0.1$, $\kappa=1$ and $\kappa=10$ cases, (heavy impurity, equal mass, and light impurity) in the upper, middle and lower panels respectively. For each $\kappa$ we have taken the initial state to be the ground state and calculated the sum in Eq. (\ref{eq:ExpectR}) over $q$, $q'$ and $s$ up to $N_{\rm max}=3,6,12$ and $24$. Similar to $r=|\vec{r}_{1}-\vec{r}_{2}|$ in the two body case \cite{kerin2020two} we observe in the backwards quench that the particle separation diverges away from $\omega t= n\pi$. For the three-body case we find that $\langle \tilde{R}[t=(n+1/2)\pi/\omega] \rangle$ diverges logarithmically with $N_{\rm max}$. More specifically we find that if the number of terms in the sums over $q$ and $q'$ is fixed then as more terms are included in the sum over $s$ $\langle \tilde{R}(t) \rangle$ converges. The divergence in $\langle \tilde{R}(t) \rangle$ comes from the sums over $q$ and $q'$ and therefore from the hyperradial wavefunction. This divergence warrants further scrutiny, to this end we investigate the evolution of the probability distribution of $R(t)$, 
\begin{eqnarray}
P(R',t)=\bra{\Psi(t)}\delta(R-R')\ket{\Psi(t)}.
\label{eq:ProbDensityDefn}
\end{eqnarray}

\begin{figure}[H]
\includegraphics[height=4.5cm, width=8.5cm]{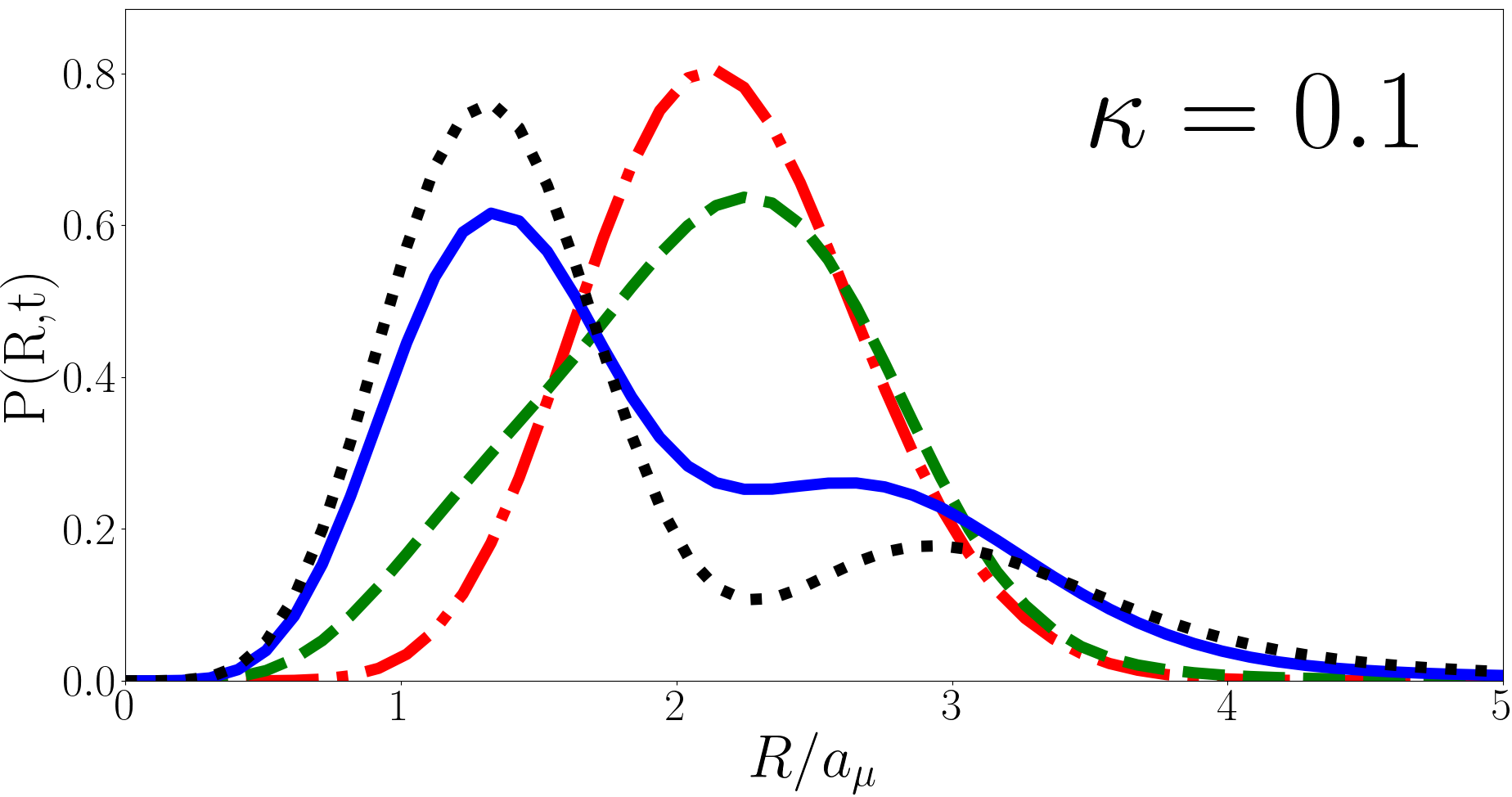}
\includegraphics[height=4.5cm, width=8.5cm]{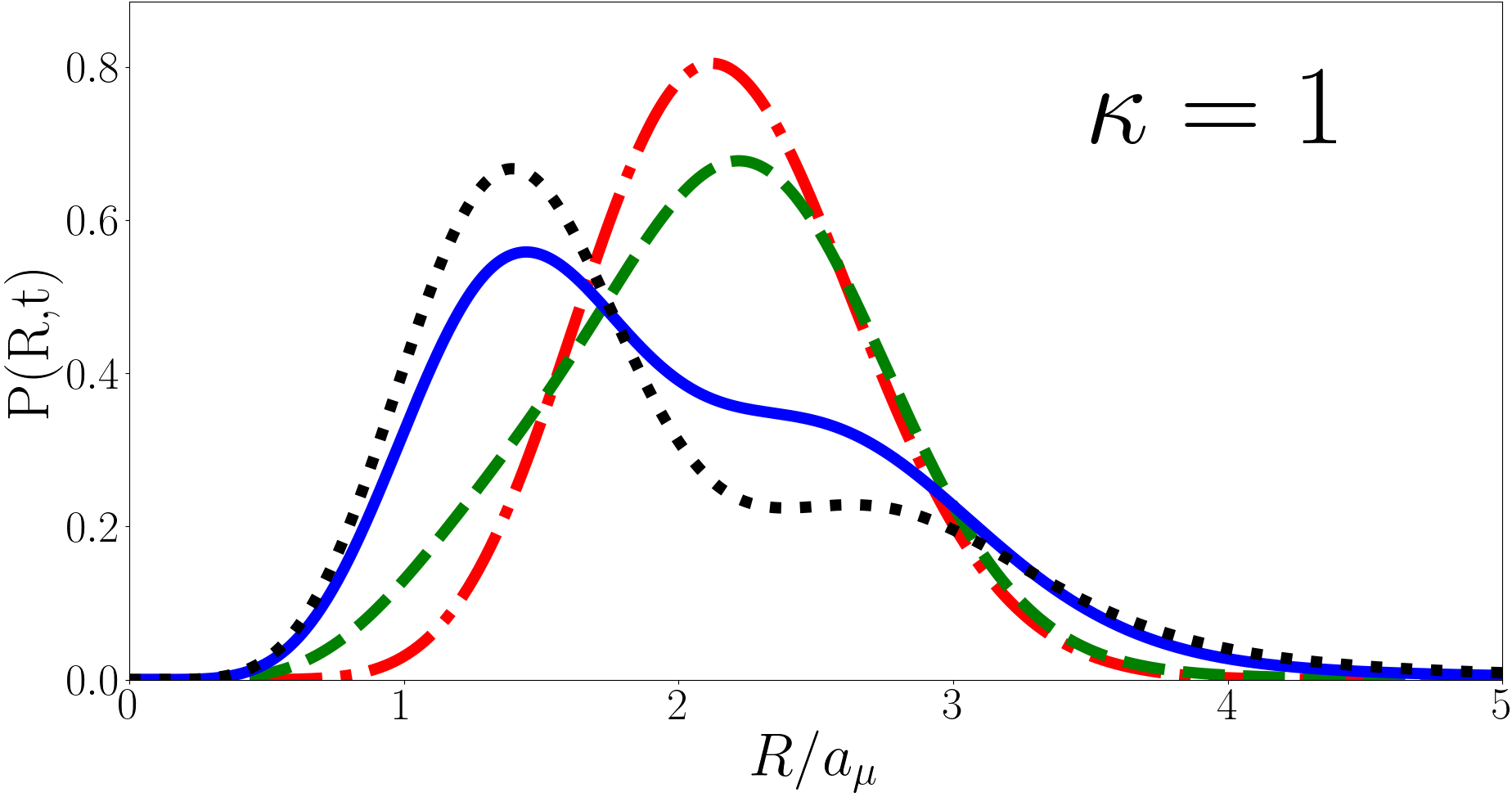}
\includegraphics[height=4.5cm, width=8.5cm]{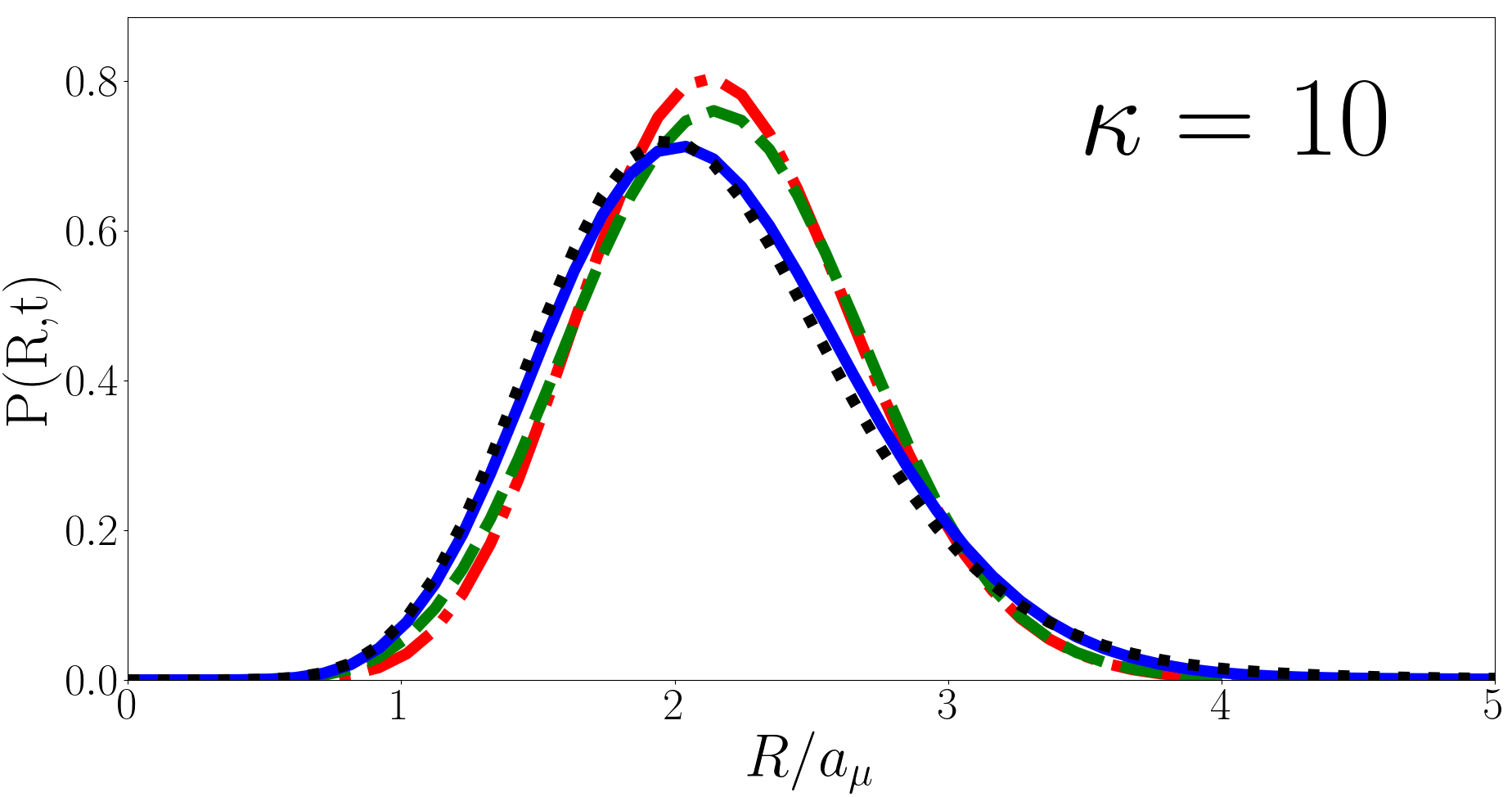}
\caption{The evolution of the hyperradial probability distribution, Eq. (\ref{eq:ProbDensityDefn}), following a forwards quench. Each plotted curve is $P(R,t)$ for a specified value of $t$ and the upper, middle and lower panels correspond to $\kappa=0.1,1$ and $10$, the heavy impurity, equal mass and light impurity cases, respectively. The red dot-dashed line corresponds to $t=0$, the dashed green line to $t=0.17\pi/\omega$, the solid blue line to $t=0.34 \pi/\omega$ and the dotted black line to $t=0.5\pi/\omega$. For all plots the initial state is $(q_{\rm i},s_{\rm i})=(0,4)$ and calculations are performed with $N_{\rm max}=24$. Each curve is convergent with $N_{\rm max}$.}
\label{fig:PRtForwardEvolution}
\end{figure}

In Fig. \ref{fig:PRtForwardEvolution} we plot $P(R,t)$ at $t=0,0.17\pi/\omega, 0.34\pi/\omega$, and $\pi/\omega$ for $\kappa=0.1,1$, and $10$ (heavy impurity, equal mass, and light impurity respectively) as a function of $R$ for the forwards quench. We find that the peak of the probability distribution shifts inwards as the system evolves, reaching its innermost point at $t=\pi/2\omega$ then evolving in reverse back to its original configuration and continuing to oscillate in this way with period $t=\pi/\omega$. The magnitude of oscillation is smaller for larger $\kappa$. This oscillatory behaviour and its dependence on $\kappa$ is unsurprising given the behaviour observed in Fig. \ref{fig:ExpectRForwards}. However the double peak, which is present for $\kappa\lesssim 5$, is unusual. We can illuminate the double peak structure by looking at the overlaps of the pre- and post-quench wavefunctions. Looking at the $\kappa=0.1$ (heavy impurity) case the largest overlaps are $|\bra{F_{0,4}\phi_{0,4}}\ket{F_{0,2.004}\phi_{0,2.004}}|^2\approx0.539$, $|\bra{F_{0,4}\phi_{0,4}}\ket{F_{0,5.817}\phi_{0,5.817}}|^2\approx0.134$ and $|\bra{F_{0,4}\phi_{0,4}}\ket{F_{1,2.004}\phi_{0,2.004}}|^2\approx0.179$. $\langle \tilde{R} \rangle$ of these states projected onto are $\approx1.66$, $\approx2.56$, and $\approx2.07$ respectively. Compare this to the $\kappa=10$ (light impurity) case where the most significant states are $|\bra{F_{0,4}\phi_{0,4}}\ket{F_{0,3.3169}\phi_{0,3.3169}}|^2\approx0.607$, and $|\bra{F_{0,4}\phi_{0,4}}\ket{F_{0,4.707}\phi_{0,4.707}}|^2\approx0.350$ and $\langle \tilde{R} \rangle$ of the states projected onto are $\approx2.01$ and $\approx2.33$. For small $\kappa$ the initial state most heavily overlaps with a few states with a relatively large variation in $\langle \tilde{R} \rangle$ for large $\kappa$ the overlaps are with states that are more closely clustered in $\langle \tilde{R} \rangle$ hence $P(R,t=\pi/2\omega)$ is singly peaked in the latter case. Physically speaking the oscillation amplitude grows smaller for a lighter impurity particle because the less mass (and therefore momenta) it has the less it is able to affect the positions of the two fermions. For the forwards quench we find $P(R,t)$ is convergent as $N_{\rm max}\rightarrow \infty$ at all points in time, as expected given $\langle R(t) \rangle$ is convergent for the forwards quench. We now turn to the divergent case, the reverse quench.

\begin{figure}[H]
\includegraphics[height=4.5cm, width=8.5cm]{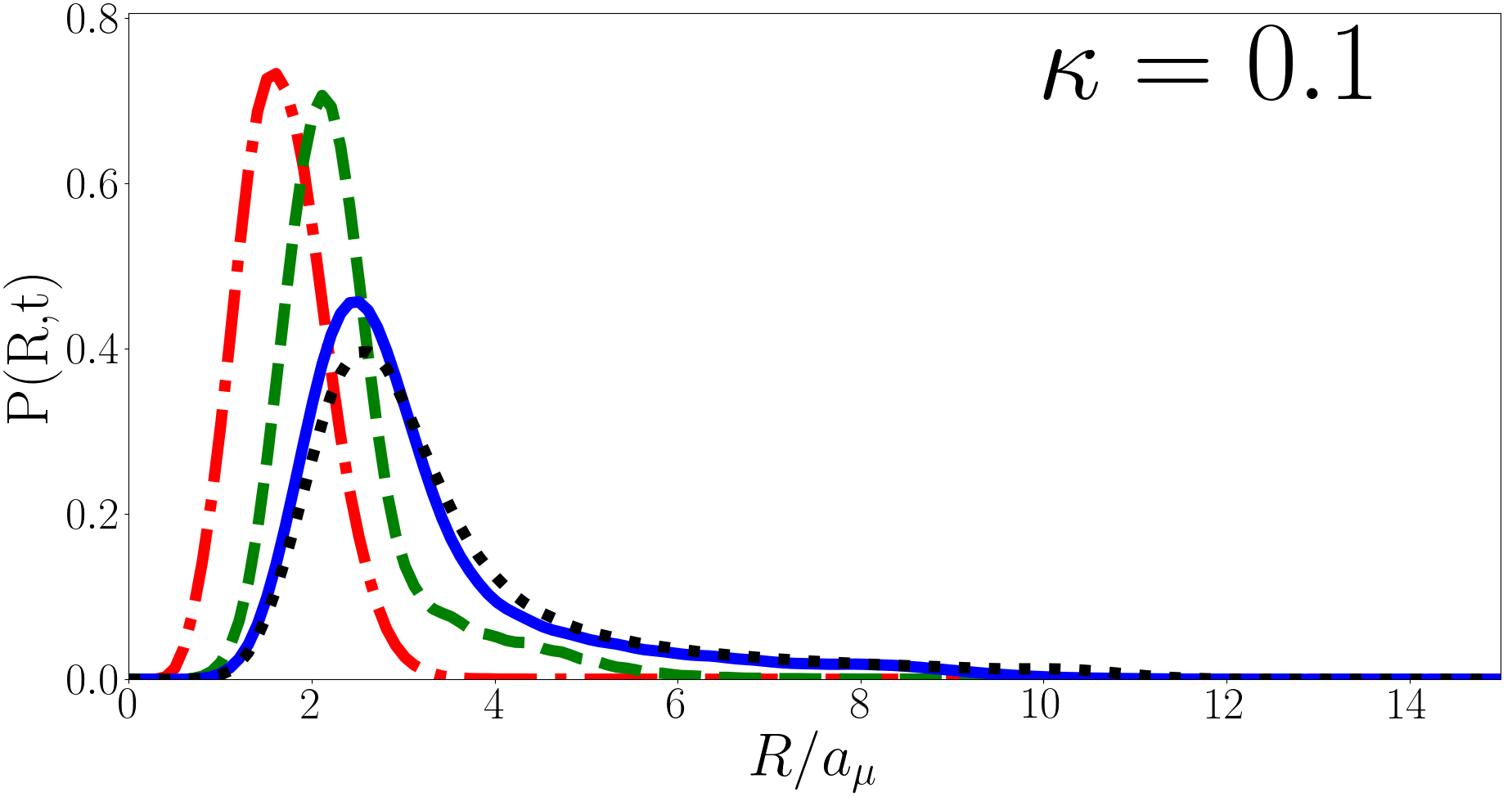}
\includegraphics[height=4.5cm, width=8.5cm]{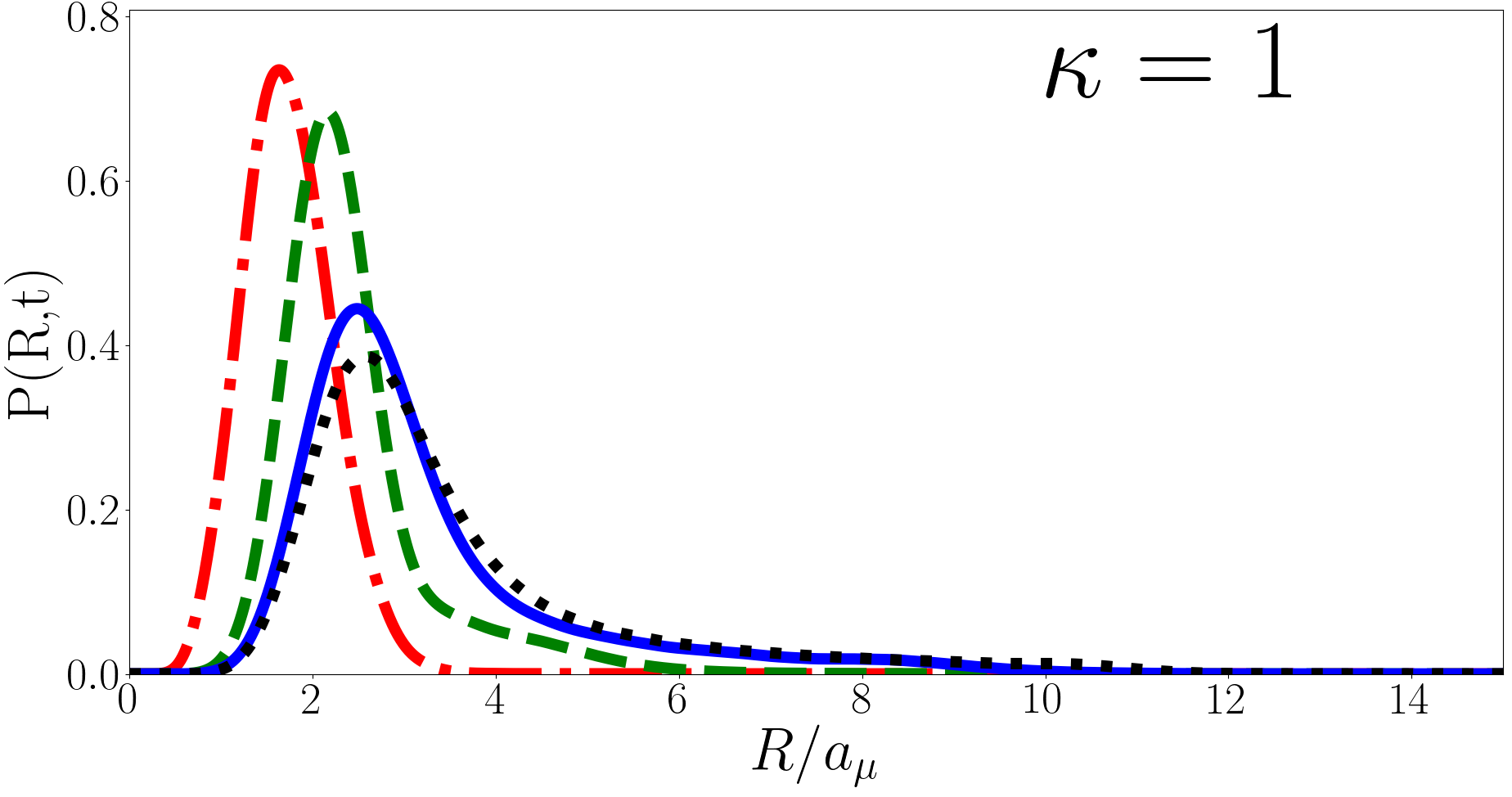}
\includegraphics[height=4.5cm, width=8.5cm]{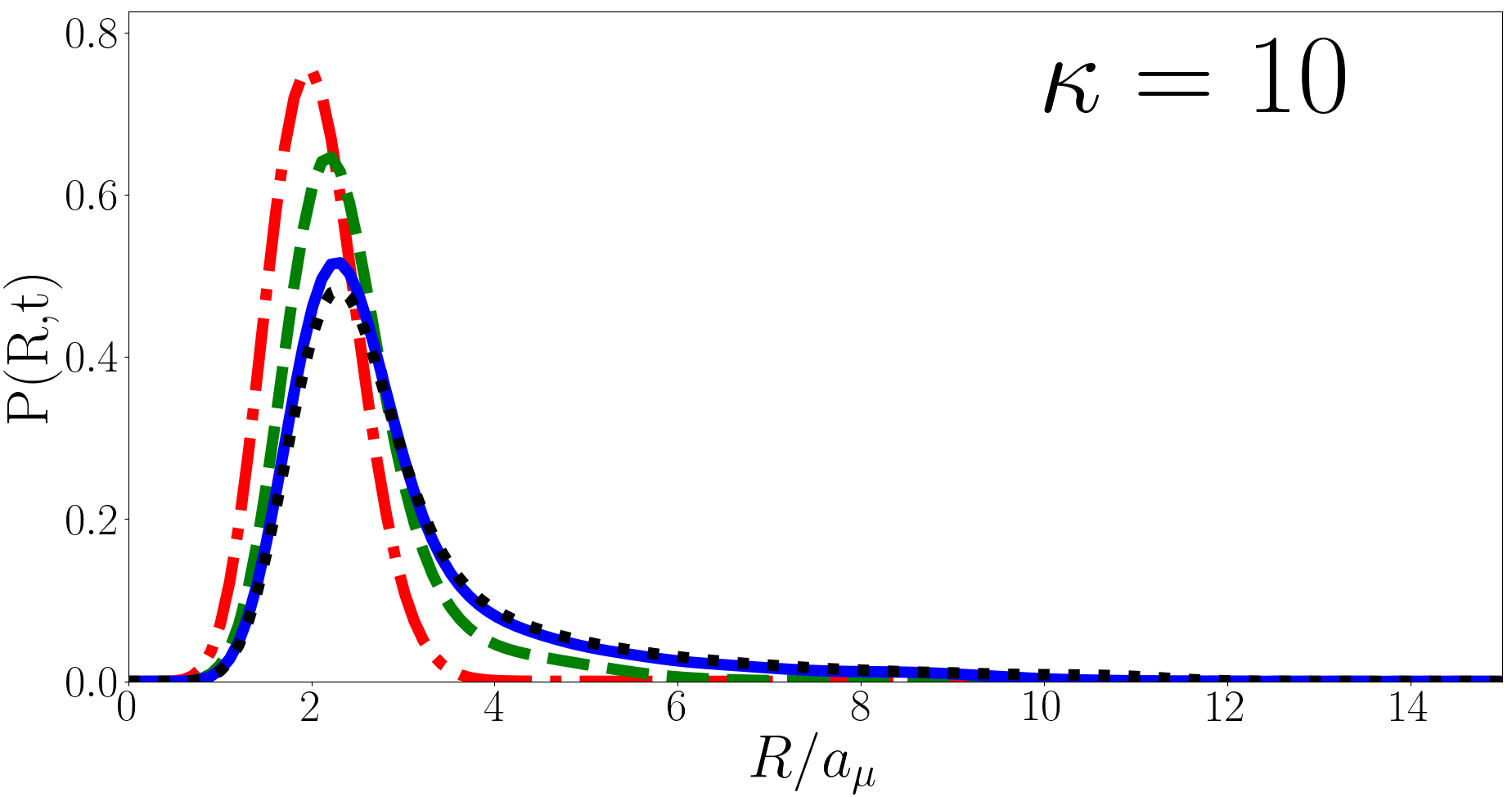}
\caption{The evolution of the hyperradial probability distribution, Eq. (\ref{eq:ProbDensityDefn}), following a backwards quench. Each plotted curve is $P(R,t)$ for a specified value of $t$ and the upper, middle and lower panels correspond to $\kappa=0.1,1$ and $10$, the heavy impurity, equal mass and light impurity cases, respectively. The red dot-dashed line corresponds to $t=0$, the dashed green line to $t=0.17\pi/\omega$, the solid blue line to $t=0.34 \pi/\omega$ and the dotted black line to $t=0.5\pi/\omega$. The initial state is given $q_{\rm i}=0$ for all plots, for the upper, middle and lower plots $s_{\rm i}=2.004\dots, 2.166\dots$ and $3.3169\dots$ respectively and all calculations are performed with $N_{\rm max}=24$. Only the $t=0$ curve (red dot-dashed) is convergent with $N_{\rm max}$. }
\label{fig:PRtBackwardEvolution}
\end{figure}

\begin{figure}[H]
\includegraphics[height=4.5cm, width=8.5cm]{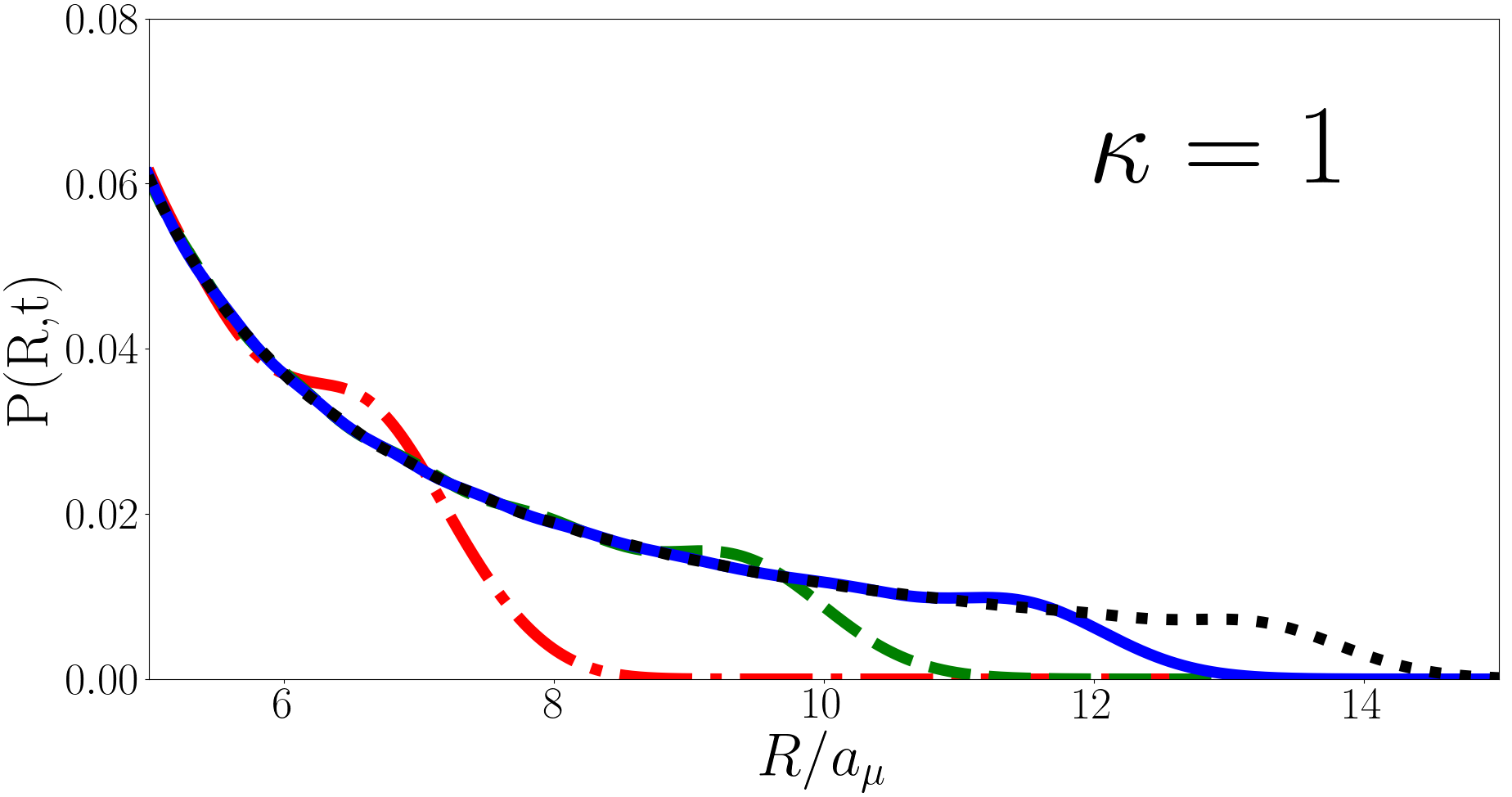}
\caption{The tail of $P(R,t=\pi/2\omega)$ in the reverse quench with $\kappa=1$ (equal mass) and $(q_{\rm i},s_{\rm i})=(0,2.166\dots)$ for various values of $N_{\rm max}$. The dot-dashed red line corresponds to $N_{\rm max}=10$, the dashed green line to $N_{\rm max}=20$, the solid blue line to $N_{\rm max}=30$, and the dotted black line to $N_{\rm max}=40$.}
\label{fig:PRtBackwardTail}
\end{figure}

In Fig. \ref{fig:PRtBackwardEvolution} we plot $P(R,t)$ at $t=0,0.17\pi/\omega, 0.34\pi/\omega$, and $\pi/\omega$ for $\kappa=0.1,1$, and $10$ (heavy impurity, equal mass, and light impurity respectively) as a function of $R$ for the backwards quench. As with the forwards quench $P(R,t)$ oscillates with period $\pi/\omega$ however the peak initially moves outwards rather than inwards. The magnitude of the oscillations grows smaller for larger $\kappa$ (lighter impurity) but the behaviour is qualitatively similar whatever the mass imbalance. For $t=n\pi/\omega$ $P(R,t)$ is approximately a Gaussian and converges with $N_{\rm max}$, however away from $t=n\pi/\omega$ the probability distribution develops a long tail. The behaviour of the tail at $t=\pi/2\omega$ for $\kappa=1$ is plotted in Fig. \ref{fig:PRtBackwardTail} for various $N_{\rm max}$ and the behaviour is qualitatively similar for different $\kappa$. The long tail decays approximately as $1/\tilde{R}^2$ before becoming exponentially suppressed, a ``cut-off''. The larger $N_{\rm max}$ is the later the cut-off occurs. While $P(R,t=\pi/2\omega)$ is normalised as $N_{\rm max}\rightarrow \infty$ this increasing long tail means that integral of $RP(R,t=\pi/2\omega)$ from $R=0$ to $R\rightarrow\infty$ is divergent, hence $\langle \tilde{R} \rangle$ diverges.

In the two-body case it has been suggested that the divergence in $\langle r \rangle=\langle |\vec{r}_{1}-\vec{r}_{2}|\rangle$ is due to the $1/r$ divergence in the two-body relative wavefunction\cite{kerin2020two}. In the three-body case the hyperradial part of Eq. (\ref{eq:Ansatz}) does not have a $1/R$ divergence and has a cusp at $\tilde{R}\rightarrow0$ for the unitary $\kappa=0.1$ (heavy impurity) and $\kappa=1$ (equal mass) ground states but not for the $\kappa=10$ (light impurity) ground state. Nonetheless the divergence in $\langle \tilde{R}(t) \rangle$ is present in all three cases. More specifically we find that for $s_{\rm i}< 3$ the initial wavefunction exhibits a cusp, whilst for $s_{\rm i}\geq 3$ there is no cusp in the initial wavefunction. Regardless of which regime the initial state is in the logarithmic divergence of $\langle \tilde{R} (t) \rangle$ persists.

However, it is clear that the finite range of the interaction in a real system provides a natural cut-off in the sum in Eq. (\ref{eq:ExpectR}). A finite range of interaction defines a minimum de Broglie wavelength which in turn defines a maximum energy which defines a maximum number of terms in the sums of Eq. (\ref{eq:ExpectR}) and thus a maximum $\langle \tilde{R} (t) \rangle$. For a system of three sodium atoms (i.e. $\kappa=1$) in a 1kHz trap and an interaction cut-off of $10^{-9}$m we obtain a cut off energy of $E_{\rm rel}\approx 8.7\times 10^6\hbar\omega$ and thus expect a maximum $\langle \tilde{R}(t) \rangle$ of $\approx 11$. With this cut-off $\langle \tilde{R}(t) \rangle$ for the backwards quench oscillates between $\approx 1.5$ and $\approx 11$, an amplitude of $\approx 5$. This is significantly larger than in the forwards quench where $\langle \tilde{R}(t) \rangle$ oscillates between $\approx1.8$ and $\approx2.2$. In light of the divergence it is natural to consider the effect of using finite-range interaction models rather than zero-range as done here, and the case of two-bodies with a soft-core interaction has been solved analytically in a one-dimensional harmonic trap \cite{koscik2018exactly}. In the limit of small interaction range it is likely that the dynamics will be similar, but the effects of longer range interactions on the dynamics are unclear. If the source of the $\langle \tilde{R}(t) \rangle$ divergence is indeed the zero-range nature of the interaction then the finite-range model may not have the divergence present in the zero-range model.

However the zero-range interaction is not the only non-physical aspect of this model. We assume that the quench in $a_{\rm s}$ is instantaneous, in experiment $a_{\rm s}$ will change continuously over some finite time. The instantaneous quench we consider here may also be related to the $\langle \tilde{R}(t) \rangle$ divergence in the backwards quench, but it is difficult to calculate a non-instantaneous quench in this formalism. In the two-body formalism it is possible to quench between any two scattering lengths \cite{kerin2020two} and thus numerically calculate a non-instantaneous quench. It would be interesting to see how this would affect the dynamics of the system and if this affects the divergence.

\section{Conclusion}
In this work we examine the time evolution of quenched systems. We consider a harmonically trapped system of two identical fermions plus a distinct particle interacting via a contact interaction where the system is quenched from non-interacting to strongly interacting or vice versa. We calculate the static wavefunction in both the non-interacting and strongly interacting regimes for general mass and use these solutions to calculate two time-dependent post-quench observables; the Ramsey signal, $S(t)$, and the particle separation, $\langle \tilde{R}(t) \rangle$. These observables are calculated for both the forwards and backwards quenches.

For the Ramsey signal we find an irregularly repeating signal for the forwards quench. This irregularity is due to the irrationality of the unitary energy spectrum and this irregular behaviour is more pronounced for small $\kappa$ (heavy impurity). For the reverse quench the magnitude of the Ramsey signal oscillates with period $\pi/\omega$ as the non-interacting energies are integer multiples and the phase has an irregular period due to the irrational initial energy.

For the particle separation the forwards quench yields the expected oscillating result, however the period is $\pi/\omega$ as the irrational contributions to the unitary eigenenergies cancel. For the backwards quench we find that the particle separation diverges logarithmically, similar to divergence in $r=|\vec{r}_{2}-\vec{r}_{1}|$ for the same quench performed on the two-body system \cite{kerin2020two}. By enforcing a cut-off based on the van der Waals interaction range we estimate a very large amplitude of oscillation, $\approx5a_{\mu}$ for $\kappa=1$. From a physical perspective it is unclear why the divergence occurs for the backwards but not forwards quench but it is likely related to the zero-range interaction and/or the instantaneous quench.

Finally we note that experimental testing of these theoretical predictions is within reach. Few-atom systems can be reliably constructed with modern techniques \cite{serwane2011deterministic, murmann2015two, zurn2013pairing, zurn2012fermionization, PhysRevLett.96.030401}, the quench in s-wave scattering length can be achieved using Feshbach resonance \cite{fano1935feshbackh, feshbach1958feshbackh, tiesinga1993feshbackh,chin2010feshbach}, and the Ramsey signal can be measured using Ramsey interferometry techniques \cite{cetina2016ultrafast}. Notably Ref. \cite{guan2019density} measured the particle separation of two harmonically trapped {}\textsuperscript{6}Li atoms following a quench in trap geometry rather than s-wave scattering length. Additionally there have been theoretical advances that allow for the four-body wavefunction to be characterised analytically in the untrapped case for 3+1 and 2+2 fermi systems \cite{endo2015absence, castin2010four}. These advances may allow for this work to be generalised to the four-body case.

\section{Acknowledgements}
A.D.K. is supported by an Australian Government Research Training Program Scholarship and by the University of Melbourne.
 
With thanks to Victor Colussi for illuminating discussions regarding the evaluation of the hyperangular integral.

\bibliographystyle{apsrev4-1}{}
\bibliography{Few-Body-Refs}

\end{document}